\documentclass{aa} % for a normal, 2 column version
\usepackage{txfonts,epsfig,graphicx,url}
\usepackage[labelfont=footnotesize,textfont=footnotesize]{caption}
\usepackage{color} % for comments and suggested changes etc.

\usepackage{natbib,twoopt}
 \usepackage[breaklinks=true]{hyperref} %% to avoid \citeads line fills
 \bibpunct{(}{)}{;}{a}{}{,}    %% natbib format like A&A and ApJ
 \newcommandtwoopt{\citeads}[3][][]{\href{http://adsabs.harvard.edu/abs/#3}%
                                        {\citealp[#1][#2]{#3}}}
 \newcommandtwoopt{\citepads}[3][][]{\href{http://adsabs.harvard.edu/abs/#3}%
                                        {\citep[#1][#2]{#3}}}
 \newcommandtwoopt{\citetads}[3][][]{\href{http://adsabs.harvard.edu/abs/#3}%
                                        {\citet[#1][#2]{#3}}}
 \newcommandtwoopt{\citeyearads}[3][][]%
   {\href{http://adsabs.harvard.edu/abs/#3}{\citeyear[#1][#2]{#3}}}

\begin{document}

   \title{Solar Fe abundance and magnetic fields}
   \subtitle{Towards a consistent reference metallicity}

   \author{D. Fabbian
          \inst{1,2}
          \and
          F. Moreno-Insertis
          \inst{1,2}
          \and
          E. Khomenko
          \inst{1,2}
          \and
          \AA. Nordlund\inst{3}
          }

   \institute{Instituto de Astrof\'isica de Canarias (IAC),\\
              Calle V\'ia L\'actea s/n, E-38200 La Laguna, Tenerife, Spain\\
              \email{[damian;khomenko;fmi]@iac.es}
             %\thanks{Also affiliated with Departamento de Astrof\'isica, Universidad de La Laguna (ULL), E-38205 La Laguna, Tenerife, Spain}
         \and
              Departamento de Astrof\'isica, Universidad de La Laguna (ULL), E-38205 La Laguna, Tenerife, Spain
         \and
              Niels Bohr Institutet, K$\o{}$benhavns Universitet,\\
              Blegdamsvej 17, DK-2100 K$\o{}$benhavn $\O{}$, Denmark\\
             \email{aake@nbi.dk}
              \thanks{Also affiliated with Centre for Star and Planet Formation (STARPLAN),
          K$\o{}$benhavns Universitet, $\O{}$ster Voldgade 5-7, DK-1350 K$\o{}$benhavn $\O{}$, Denmark}
             }

   \date{Received April 03, 2012; accepted August 20, 2012}

% \abstract{}{}{}{}{}
% 5 {} token are mandatory

  \abstract
   {}
   {We investigate the impact on Fe abundance
   determination of including magnetic flux in series of 3D radiation-MHD
   simulations of solar convection which we used to synthesize spectral intensity profiles corresponding to disc centre.}
   {A differential approach is used to quantify the changes in
     theoretical equivalent width of a set of $28$ iron spectral lines
     spanning a wide range in wavelength, excitation potential,
     oscillator strength, Land\'e factor, and formation height.
     The lines were computed in LTE using the spectral synthesis code
     {\texttt{LILIA}}. We used input magnetoconvection
     snapshots covering 50 minutes of solar evolution and belonging to
     series having an average vertical magnetic flux density of
     $\langle {\mathrm B_{vert}} \rangle = 0, 50,
     100,~\mathrm{and}~200$ G. For the relevant calculations we used
     the Copenhagen {\texttt{Stagger}} code.}
{The presence of magnetic fields causes both
   a direct (Zeeman-broadening) effect on spectral lines with non-zero
   Land\'e factor and an indirect effect on temperature-sensitive lines via a
   change in the photospheric $T-\tau$ stratification. The corresponding
   correction in the estimated atomic abundance ranges from a few hundredths
   of a dex up to $|\Delta \mathrm{\log} \, \epsilon \mathrm{(Fe)}_{\odot}|
   \sim 0.15$\, dex, depending on the spectral line and on the amount of
   average magnetic flux within the range of
   values we considered. The Zeeman-broadening effect gains relatively more
   importance in the IR. The largest modification to previous solar abundance
   determinations based on visible spectral lines is instead due to the
   indirect effect, i.e., the line-weakening caused by a warmer
   stratification as seen on an optical depth scale. Our results indicate
   that the average solar iron abundance obtained when using
   magnetoconvection models can be $\sim 0.03-0.11$\, dex higher than when
   using the simpler HD convection approach.}
   {We demonstrate that accounting for magnetic flux is important in
     state-of-the-art solar photospheric abundance determinations
     based on 3D convection simulations.}


   \titlerunning{Solar Fe abundance and magnetic fields}
   \authorrunning{D. Fabbian et al.}

   \maketitle
%
%________________________________________________________________

\section{Introduction}
\label{intro}

The chemical composition of stars is derived through the analysis of
their emitted electromagnetic radiation.  Thanks to appropriate
modelling, the presence of lines of a given chemical element in the
observed spectra can be interpreted in terms of its abundance in the 
source object.  The line strength, as quantified, say, by the equivalent
width, needs to be matched by the corresponding theoretical prediction. A
more detailed approach involves matching the line profiles themselves
instead.

As our closest star, the Sun allows the most detailed observations
of the processes taking place in an object of its type. The
atmospheric models used to describe the physical conditions of the
solar plasma have become more and more realistic through the years, in
particular, accounting for 3D inhomogeneities and for the need to have
good granulation statistics.  Still, magnetic fields are usually
neglected in solar abundance studies, on the assumption that their
effect on spectral line formation is of secondary importance. The
space- and time-averaged solar spectrum commonly used for abundance
determinations will in fact be weighted towards granules, owing to the
larger surface area the latter cover with respect to
  intergranular lanes, where magnetic field concentrations tend to be
  found. The idea is that for this reason the impact of magnetic flux
over the average emergent spectrum (and, in particular,
  abundances) is likely minimized. According to this assumption, the
uncertainty introduced by using magnetic-free models would be less
important than other more obvious sources of error, such as atomic
data, uncertainties related to observational data, treatment of
non--LTE effects, and so on. Moreover, the assumption that magnetic
fields could be disregarded altogether in abundance studies has been
compounded with the need of simplifying the simulation setup, spectral
synthesis computations and subsequent analysis. Because of to this situation,
when advancing from the 1D plane-parallel approximation to a 3D
approach, non-magnetic model atmospheres were used. Nevertheless, two
major arguments can be put forth against that simplification.

1. From an observational point of view, evidence is accumulating
about a large amount of hidden magnetic flux being present in the Sun,
even in ``quiet'' regions (see \citeads{2011ASPC..437..451S} for a recent review).
It is now known that weak magnetic fields cover most of the Sun's surface, with
indications that a flux density
of order $10^2$ G may be a good reference value
(\citeads{2004Natur.430..326T}; \citeads{2009LRSP....6....2N}).
In regions with strong solar
magnetic fields (e.g., in plage regions),
the fraction of magnetic field concentrations with
of the order of 1 kG in intergranular lanes tends to increase (see
e.g. Fig. 46 in \citeads{2009LRSP....6....2N}).
In observed plage regions, the granulation presents a lot of fine
structuring at high spatial resolution: isolated bright points,
strings of bright points and dark micro-pores, ``ribbons'', or more
circular ``flower'' structures \citep{1992ApJ...393..782T,
2004A&A...428..613B, 2010A&A...524A...3N}. Granules become smaller
than in the quiet areas \citep{1977SoPh...52..249M,
1988A&A...197..306S, 1992ApJ...393..782T}, and intergranular lanes are
characterized by the presence of micro-pores.
In agreement with observations, numerical 3D convection models predict that
granulation pattern is profoundly affected in regions with high flux density
(e.~g. \citeads{CattEmonWeiss2003};\citeads{Vogler2005},
\citeads{2008ApJ...687.1373C}), with reduced size of the upflowing granules
and widened intergranular lanes. In the extreme case of sunspot umbrae, the
convection is seriously impaired and takes the form of narrow upflow plumes
(\citeads{2006ApJ...641L..73S}).

2. The reasoning given for why HD should be a good
approximation for line formation in the Sun neglects the fact that the
photospheric material becomes more transparent in magnetic
concentrations owing to their lower density, allowing one
to probe deeper, i.e.,  into hotter layers of the atmosphere
(\citeads{2000SoPh..192...91S}) where the flux tubes are also heated through
heat influx from the surrounding material (\citeads{Spruit1976}).
Weak spectral lines will then experience brightening of their core.
The weakening of neutral iron spectral lines
was, for example, used by \citetads{SchussSolanki1988} to predict a
higher continuum intensity for magnetic elements. Observed
photospheric bright points may then be identified with magnetic flux
concentrations.
Magnetic fields can therefore act on
temperature-sensitive spectral lines not only directly, i.e.,
via Zeeman broadening, but also via an {\it indirect} effect associated with
the temperature stratification in the line-forming regions.
Only a comprehensive study using realistic 3D magnetohydrodynamical
simulations can therefore confirm whether the effect of magnetic fields on
spectral line formation and abundance derivation is truly negligible.

Including magnetic fields is clearly an important new step in the
long-term evolution of abundance studies. Following pioneering work (e.g.
\citeads{Nordlund1984}), 3D radiative magnetoconvection
simulations that explore realistic configurations and provide an excellent
match to the photometric and spectroscopic observations were produced during the past decade, as summarized by \citetads{2009LRSP....6....2N}.  The
resulting 3D model atmospheres (albeit with zero magnetic field) have been
used in abundance determination work to realistically account for the
departures from homogeneity in line formation. Along with the use of better
input atomic data, these advances led to remarkable agreement between the
observational data and the theoretically predicted shapes and asymmetries of
the spectral lines (\citeads{2000A&A...359..729A, 2001ApJ...550..970S}), but
only if decreased abundance of several chemical elements in the Sun was
assumed (\citeads{2000A&A...359..729A}; \citeads{2005ARA&A..43..481A};
\citeads{2009ARA&A..47..481A}) compared to the previously commonly adopted
standard reference values inferred via a 1D static, homogeneous, and
plane-parallel LTE analysis.  The 3D-derived abundances give smaller
line-to-line scatter compared to the 1D results at the same time as rendering the microturbulence and macroturbulence free parameters unnecessary. Better agreement with solar neighbourhood abundances
\citepads{2009LanB...4B...44L} is also obtained.

While all these results are very encouraging, pressing
outstanding issues remain. Of particular importance among them is the neglect of
magnetic fields in the convection models mentioned above, an issue that we approach
in this paper. The role of magnetic fields in abundance determinations has
been initially investigated in 1D model atmospheres \citepads{Borrero2008},
finding non--negligible effects, but it was not until very recently
(\citeads{2010ApJ...724.1536F}) that an investigation set out to explore the
effects in the 3D MHD case. The results of that exploratory paper indicated
that, even in the magnetically quiet Sun, the neglect of magnetic fields may
question the few-centidexes accuracy claimed in previous works.

In the present paper, we achieve the first comprehensive 3D study
  of the solar iron abundance that includes magnetic fields. We focus
on iron because of the large number of spectral lines for this element
in the visible and IR with different direct magnetic sensitivity
(i.e., differing Land\'e factors g$_{\mathrm{L}}$) and also because
the solar Fe abundance is the basis for most abundance work in the
astrophysics literature.  Iron is in fact of central historical
importance as a proxy for the total metal content of a given stellar
object and as the reference element in the elemental abundance ratios
used in studies of galactic chemical evolution.  The Fe solar
abundance was claimed in the 80's to lie anywhere between
$\mathrm{\log} \, \epsilon \mathrm{(Fe)}_{\odot} = 7.70$\, dex\,
(\citeads{1984A&A...132..236B}, \citeyearads{1986MNRAS.220..549B}) and
$\sim 7.48$\, dex\, \citepads{1990A&A...232..510H}.  The use of 3D
model atmospheres and improved observational, atomic, and modelling
data led to a drastic reduction of the large line-to-line abundance
scatter, whereby a low abundance of $\mathrm{\log} \, \epsilon
\mathrm{(Fe)}_{\odot} = 7.45$\, dex \citepads{2000A&A...359..743A} was
determined, consistent with the meteoritic value of
\citetads{2009LanB...4B...44L}.  The question remains whether that
mean value can withstand the inclusion of magnetic effects in the 3D
model atmospheres by using {\it magneto}convection (as opposed to
purely HD) models.

In this paper we find
confirmation of the importance of magnetic fields in determining the
solar Fe abundance. For the Fe spectral lines we have
in common with \citetads{2000A&A...359..743A},
we find abundance corrections of $\Delta \mathrm{\log}
\, \epsilon \mathrm{(Fe)}_{\odot}=0.05-0.09$\, dex when considering
magnetoconvection models with an average unsigned magnetic flux of $\langle {\mathrm B_{vert}} \rangle = 100$ G.

The neglect of magnetic fields is not the only remaining problem in the
abundance determination research field. Other open questions are (i) the friction
with helioseismic inversion constraints (\citeads{2005ApJ...620L.129A};
\citeads{2006ApJ...649..529D}) caused by the up to $\sim 30 \%$ smaller
estimate currently widely adopted for the solar photospheric metal content; (ii) several authors (e.g.,
\citeads{2007A&A...473L...9C}; \citeads{2007AAS...211.5908A};
\citeads{2008A&A...488.1031C}; \citeads{2008ApJ...686..731A};
\citeads{2010A&A...514A..92C}; \citeads{2011SoPh..268..255C}) have suggested
that the revision in the abundance of a number of chemical elements may be
too drastic, for example due to slightly too steep temperature gradients in
the employed models' continuum-forming layers (\citeads{2009ApJ...694.1364T})
or too-low temperatures in their middle photospheric layers
(\citeads{2012AAS...21914408A}); (iii) \citetads{2009MmSAI..80..643C} claim
that the revised solar abundances are likely mostly the result of differences
in equivalent width and/or line profile fitting, as well as in atomic data and
blends, rather than granulation effects and hydrodynamical modelling.

The layout of the paper is as follows.
Section~\ref{section:computations} deals with the details of the
atmospheric models employed in this study and with the method
used in computing theoretical spectral line
profiles. Section~\ref{section:abucorr} presents the abundance
correction results that we find for the different Fe lines studied here.
In Sect.~\ref{section:consequences} we draw
consequences specifically with respect to the
solar iron content. Section~\ref{section:discussion} discusses the more
general implications in terms of derivation of the elemental
composition of stars, and Sect.~\ref{section:conclusions} lists our conclusions.

%__________________________________________________________________

\section{Atmospheric models and computation of spectral lines}
\label{section:computations}

\subsection{Method}

We used simulations covering several hours of solar time that we obtained by using the 3D radiative-hydrodynamics Copenhagen
{\texttt{STAGGER}} code (\citeads{1996JGR...10113445G}, also, see discussion in \citeads{2010ApJ...724.1536F} and the recent description in \citeads{2012A&A...539A.121B}).
The simulation box covers $6.0 \mathrm{x} 6.0$\,Mm$^2$ horizontally and extends in height for $\sim 2.8$~Mm. These datacubes
are divided into $252 \mathrm{x} 126  \mathrm{x} 252$ cells, with non--uniform spacing in
the vertical direction, yielding a resolution in height of $\lessapprox 15$\,km
at the photosphere.
In the vertical direction, after avoiding the vertical ghost zones necessary in the computations,
only $\sim 2.5$~Mm actually contain valuable physics information.
Horizontally periodic boundary
conditions are imposed for all variables on the sides of the
simulation box.

The initial setup configuration for the magneto-convection simulations
is a uniform vertical magnetic field. While diverging convective
  upflows will tend to sweep and concentrate magnetic fields to
  intergranular downflow lanes (e.g., \citeads{2009LRSP....6....2N}), the
  horizontal average of the vertical component of B must remain equal
  at all depths and times throughout the simulation to the initial
  value chosen for it in the relevant series, $<B_{vert}> = B_{t=0} =$
  const, and thus characterizes the different magnetic series in this
  study. We subdivide our results into series for initial vertical
field of $0$ G (the HD series) and $50$, $100$, and $200$ G for the MHD
series.  For each series we wait until a statistically stationary
state has been reached, which typically takes some $0.5$ solar hours
after introduction of the initial vertical magnetic field. We then
store snapshots of the subsequent temporal evolution at intervals of $30$
s solar time.

\begin{figure}
   \centering
   \includegraphics[width=0.45\textwidth]{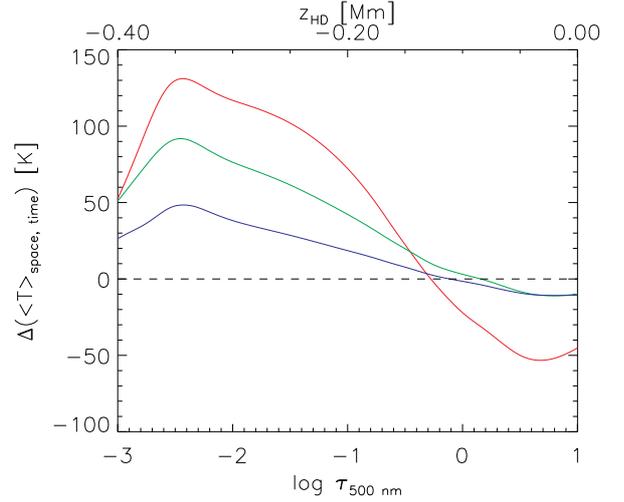}
   \caption{MHD-HD difference of the temperature stratification obtained for layers close to photospheric line-forming regions after averaging horizontally on surfaces of equal optical depth and temporally over the ($63 \mathrm{x} 63$ column) input snapshots used in the spectral
     synthesis. Blue, green and red lines represent the difference in temperature with respect to the HD case, against $\log_{10} \tau_{(500\,\mathrm{nm})}$ (bottom horizontal axis), for the 50 G, 100 G and 200 G cases, respectively. The top
     horizontal axis shows the corresponding physical depth in Mm in the HD case.}
         \label{fig1_avTxyt_vsDepth_and_vsTau_ks1in5}
   \end{figure}

For the a-posteriori spectral synthesis we used the LILIA code
(\citeads{2001ASPC..236..487S}). A subdomain of the computed box
extending from $425$\,km above to $475$\,km below the average
$\tau_{500\,\mathrm{nm}}=1$ level was used for the synthesis; the
spatial resolution in that domain was increased by interpolating to a
uniformly spaced grid with resolution $\sim 7.8$\,km. In
Fig.~\ref{fig1_avTxyt_vsDepth_and_vsTau_ks1in5} we show the MHD-HD
differences in the spatially and temporally averaged temperature
stratification.  The average temperature stratification is different
for the different series, with a tendency in layers with $\log
  \tau_{500\,\mathrm{nm}}<0$ to an increasingly large departure from the
  HD temperature stratification as the initial magnetic field becomes
  higher, following the expected behaviour (see Sect.~\ref{intro}). We
  stress that this effect appears due to the changed opacity caused by
  the decreased pressure (lower density) inside strong magnetic flux
  concentrations (magnetic plus gas pressure being nearly equal to the
  gas pressure in the surrounding unmagnetized plasma, e.g.
  \citeads{2009LRSP....6....2N}). The photospheric temperature
  increase in MHD is therefore not related to the magnetic energy
  build-up by photospheric motions, which has been hypothesized to heat
  the corona after local release via magnetic reconnection.

The synthetic spectra were obtained only for solar disc centre.
After verifying that the influence on the correspondingly derived
theoretical equivalent widths was negligible, for the spectral synthesis we
decreased the horizontal resolution by taking one out of every four grid
points. We selected a subset of snapshots covering the last 50 minutes
of the statistically stationary regime of convection, separated by a
150 solar seconds time interval.  We therefore keep $21$ snapshots per
series.

Spatially- and temporally-averaged synthetic spectra were finally
  computed and then employed for a differential analysis of the impact
  of different amounts of magnetic flux on abundance determinations.

%                                                One column figure
%-----------------------------------------------------------
%
%______________________________________________________________

\subsection{Line parameters}

\begin{table}
  \caption{Atomic parameters for the $28$ iron spectral lines included in the current study.}
\label{lines_params}      % is used to refer this table in the text
\centering                          % used for centering table
\begin{tabular}{c c c c c}          % centered columns (5 columns)
\hline\hline                 % inserts double horizontal lines
$\lambda$                  & $\chi_{l}$ & $\log gf$ & g$_{\mathrm{L}}$ & ID \\ % & H$_{\mathrm{D}}$ & H$_{\mathrm{W}}$\\    % table heading
\protect   [nm]            &    [eV]    &           &                  &   \\ % & [km]             &    [km]    \\
\hline                        % inserts single horizontal line
  410.091                &  2.4530    & $-4.244$ &  0.0                & 0L \\ % &  -  &  -  \\      % inserting body of the table
  415.025                &  3.4300    & $-1.260$ &  0.0                & 0L \\ % &  -  &  -  \\
  423.027                &  4.2310    & $-2.440$ &  0.0                & 0L \\ % &  -  &  -  \\
  430.603                &  4.2180    & $-1.922$ &  0.0                & 0L \\ % &  -  &  -  \\
  437.113                &  4.2600    & $-3.306$ &  0.0                & 0L \\ % &  -  &  -  \\
  486.364                &  3.4300    & $-1.663$ &  0.0                & 0L \\ % &  -  &  -  \\
  524.705 &  0.0873    & $-4.946$ &  2.0                & A \\ % & 415 & 328 \\ % VALD g_L=1.99; this line and the next one: good magnetic pair of Stenflo, see notes in LINES file
  525.021 & 0.1213     & $-4.938$ &  3.0                & A \\ % & 409 & 324 \\ % also: IMaX line
  525.065                & 2.1979     & $-2.181$ &  1.5                & LL \\ % & 493 & 330 \\ % enters IMaX wavelength window (see Fig. 15 in Mart\'inez Pillet et al. (2011)) as a bonus, given that it is the previous one (525.02 nm) which was the target of the IMaX experiment.
%  538.981                &  4.4460    & $-3.567$ &  3.12              & LL \\ % &     & \\ % TOO WEAK
  543.452                &  1.0110    & $-2.122$ &  0.0                & 0L \\ % &  -  &  -  \\ % VALD g_L=-0.01; Bello Gonz\'alez et al. (2010); non--LTE?
  557.609                &  3.4302    & $-0.940$ &  0.0                & 0L \\ % &  -  &  -  \\ % VALD g_L=-0.01; Bello Gonz\'alez et al. (2009) ; close to [O I] 557.7 nm
  579.119                &  5.0200    & $-2.067$ &  0.0                & 0L \\ % &  -  &  -  \\
  585.923                &  4.3010    & $-3.076$ &  0.0                & 0L \\ % & 135 & 126 \\
  594.036                &  5.0100    & $-2.260$ &  0.0                & 0L \\ % &  -  &  -  \\
  601.392                &  5.1000    & $-2.133$ &  0.0                & 0L \\ % &  -  &  -  \\
  608.271 &  2.2230    & $-3.572$ &  2.0                & A \\ % & 229 & 203 \\
  611.332 & 3.2210 & $-4.230$ &  0.57             & II \\ % & 116 & 106 \\  % Fe II line
  616.946                &  3.4170    & $-4.009$ &  0.0                & 0L \\ % &  -  &  -  \\
  617.334 & 2.2227     & $-2.880$ &  2.5                & A \\ % & 361 & 279 \\
  624.065 & 2.2230     & $-3.390$ &  1.0                & A \\ % & 276 & 231 \\ % VALD g_L=0.99
  630.150                &  3.6540    & $-0.718$ &  1.67               & LL \\ % & 489 & 286 \\ % close to [O I] 630.060 nm
  630.249                &  3.6860    & $-0.973$ &  2.5                & LL \\ % & 381 & 264 \\ % VALD g_L=2.49; % close to [O I] 630.060 nm; HINODE Solar Optical Telescope (SOT) vector spectropolarimeter (SP, Lites et al. 2001) line, solar photosphere magnetic fields measurements with a spatial resolution of 0.2 arcseconds
  633.084                &  4.7330    & $-1.740$ &  1.23               & LL \\ % & 180 & 150 \\
 1558.826                &  6.4740    & $ 0.391$ &  1.5                & IR \\ % &  -  &  -  \\
 1564.851                &  5.4260    & $-0.670$ &  2.98               & IR \\ % & - & - \\
 1565.287                &  6.2461    & $-0.170$ &  1.55               & IR \\ % & - & - \\
 1566.202                &  5.8280    & $ 0.382$ &  1.47               & IR \\ % & - & - \\
 1566.524                &  5.9790    & $-0.338$ &  0.67               & IR \\ % & - & - \\
\hline                                   %inserts single line
\end{tabular}
\tablefoot{The columns list, respectively, the rest wavelength, lower transition level excitation potential, oscillator strength, Land\'e factor of the iron spectral lines studied here, and the identifiers employed in Sect.~\ref{section:abucorr} for a useful subdivision into different groups of lines.\\
}
\end{table}
%
%_____________________________________________________________

The $28$ iron lines we chose for the spectral synthesis are listed
(with their parameters) in Table~\ref{lines_params}.
The spectral lines cover a wide range in rest wavelength ($\lambda$), with
$23$ of them in the visible and $5$ of them in the near-IR spectral
region. To detect possible dependences of the final abundance
corrections on the atomic parameters, the spectral lines were chosen so as
to include a wide range of values of lower transition level excitation
potential ($0.1 < \chi_{l} < 6.5$~eV), of transition oscillator
strength ($-5.0 < \log gf < 0.4 $), and of Land\'e factor ($0
\lesssim g_{\mathrm{L}} \lesssim 3$). The values adopted are
from the VALD (Version 0.4) and NIST ASD (Version 4)
databases\footnote{The Vienna Atomic Line Database (VALD) is available
as an online interface at
http://vald.astro.univie.ac.at/$\sim$vald/php/vald.php; the National
Institute of Standards and Technology Atomic Spectra Database (NIST
ASD) is accessible at http://www.nist.gov/pml/data/asd.cfm}.

In Table~\ref{lines_params}, we assign different identifiers for
  subgroups of lines, to reflect the organisation that will be used in the discussion of results (Sect.~\ref{section:abucorr}).
  We make a division into Zeeman-insensitive (g$_{\mathrm{L}}=0$)
  lines (Sect.~\ref{sec:gL_eq0}) - identified in the table with
  the label ``0L'' - and in lines with non-zero Land\'e factor
  (Sect.~\ref{sec:gL_neq0}). The latter are additionally
  subdivided into lines with high values of the Land\'e factor
  (``LL'' in Table~\ref{lines_params}), i.~e., having g$_{\mathrm{L}} \ge 1$
  (see Sect.~\ref{sec:lines_large_lande}), lines in common with
  \citetads{2000A&A...359..743A} (Sect.~\ref{sec:lines_common_with_asplund}) - marked in the table with
  ``A'' - and IR lines (\ref{sec:IR_lines}), identified with ``IR''.
  The spectral features listed in Table~\ref{lines_params}
  are all lines of neutral iron, except for the singly-ionized
  $611.332$~nm iron line, marked in the table with ``II''.
  The grouping just discussed will help to more clearly explain the direct
  and indirect effects of the magnetic flux on
  spectral lines.

The iron lines chosen for this study probe quite different depth
ranges in 
the photospheric snapshots, thus jointly covering a domain from the
high layers in the simulation box to as deep as
around $5$~km above $\log_{10} \tau_{(500 nm)}=0$.
Their approximate heights of formation (e.~g. see values in
\citeads[][Table~2]{1989GurtKost}) make them generally safe for this
study since they are supposed to be forming inside the simulation domain.

The collisional Van der Waals broadening/damping formula was used
following the classic approximation of \citetads{1955QB461.U55......}
with no enhancement. Our tests revealed that, while the exact choice
of the enhancement factor will influence the resulting equivalent
width and is thus an important choice for line profile fitting and
absolute abundance matching, it does not affect a {\it differential}
abundance investigation, like ours, to a significant level. For
instance, the abundance correction derived from the comparison of the
HD and MHD equivalent widths becomes less severe by less than $0.01$\, dex\, in absolute value when the enhancement factor is changed from 1.0
to 2.5.

The lines were synthesized at a spectral
resolution of 0.50 pm. We found this to be a good compromise for achieving high spectral resolution at an acceptable computational
effort with a resulting equivalent width accuracy better than $0.001$ pm.
For each spectral line we used as many spectral points around the central
wavelength as needed to reach convergence at least at the
$10^{-2}$ level when calculating the equivalent width.
For the strongest lines this implied using up to $1000$ wavelength points (or
$\sim 250$~pm on each side of the rest wavelength).

\subsection{Continuum intensity}

%______________________________________________
   \begin{figure}
   \centering
   \includegraphics[width=0.45\textwidth]{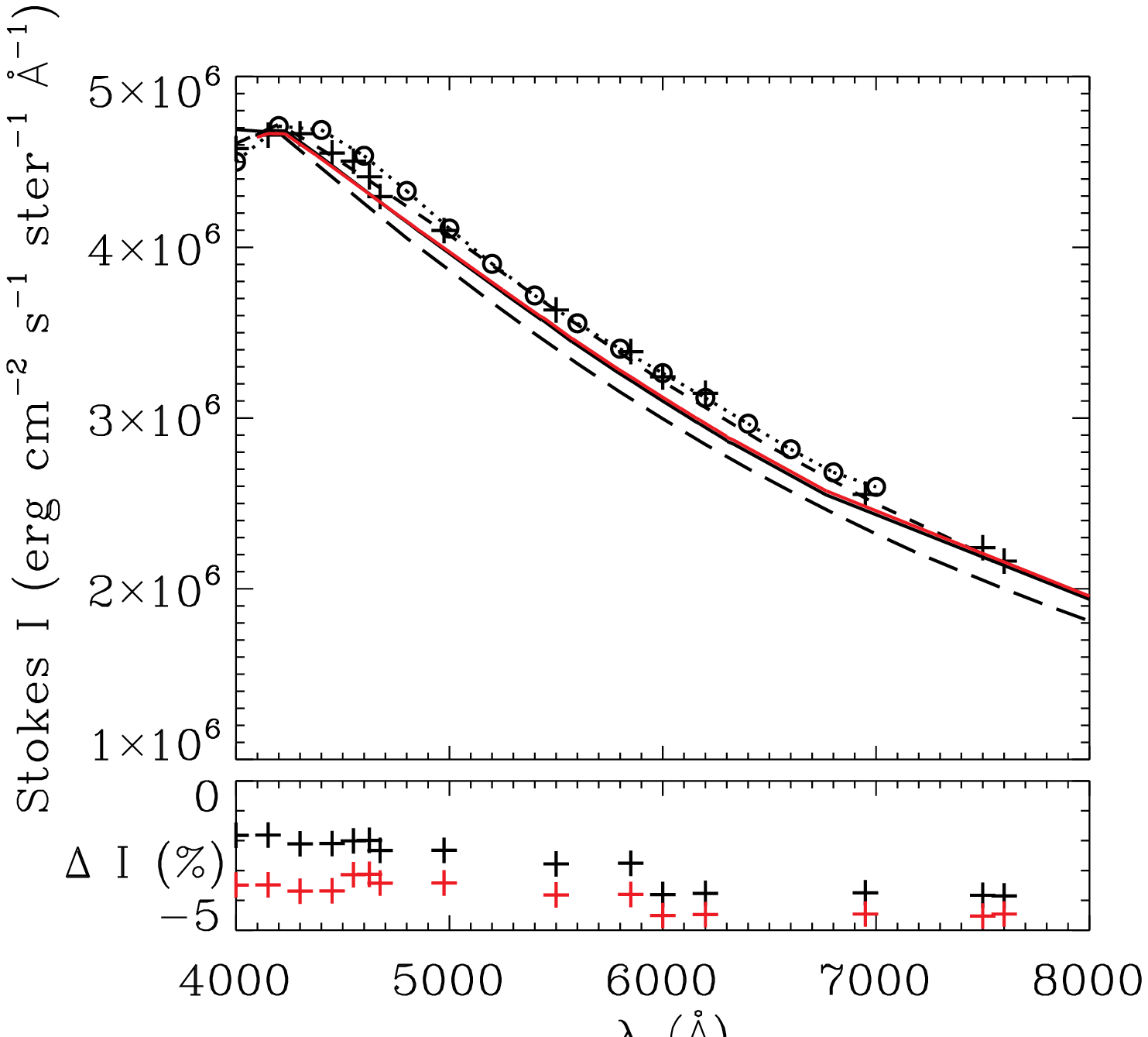}
   \caption{{\it Top panel:} comparison of our computed solar disc
     centre continuum intensity in the non-magnetic and 200 G
       case (solid black and red curves, respectively) to data
     from the literature.  The long-dashed curve is based on the
     results obtained by \citetads{2009ApJ...694.1364T} using a single
     3D HD snapshot by \citetads{2000A&A...359..729A}; the dotted
       curve with circles is based on \citetads{2006ApJS..165..618A};
     the short-dashed curve refers to the semi-empirical 1D
     MACKKL model of the ``quiet'' Sun atmosphere (see Table II of
     \citeads{1986ApJ...306..284M}); finally, black crosses indicate
     the observational data from \citetads[][Table
     VII]{1984SoPh...90..205N}. {\it Bottom panel:} intensity
       difference (in percentage) between our results and the
       observational data (black crosses=HD-obs; red crosses=200
       G-obs).}
   \label{contint}
   \end{figure}
%______________________________________________

   Figure~\ref{contint} ({\it top panel}) shows the level of
   agreement between our calculated continuum intensity data
     (based on a space and time average for the same snapshots of the
     HD series that were employed in the spectral synthesis) and
   different theoretical and measured values, all being {\it
     spatially-averaged means} at solar disc centre ($\mu = 1$) and
   given in absolute units.  Our continuum intensity results 
     for the HD and 200 G series are seen to lie between the 3D HD values that
   \citetads{2009ApJ...694.1364T} derived based on a single snapshot
   from \citetads{2000A&A...359..729A}, and
   the values from the solar photospheric ``thermal profiling''
   analysis of \citetads[][see their Table 1]{2006ApJS..165..618A}. 
   Our model lies in a band of
   $\lesssim 5$\% below the semiempirical model of
   \citetads{1986ApJ...306..284M}, known as MACKKL (short-dashed curve
   in Fig.~\ref{contint}), which was purposely built at the time
   precisely from continuum observations. Therefore, that our
   results are, in particular, close to the latter, confirms the
   goodness of our 3D MHD simulations.  They also agree well with
     the observational data from \citetads[][Table
     VII]{1984SoPh...90..205N}.  And with the continuum distribution of
     \citetads{2012KPCB...28...49S} (not shown in our figure to avoid
     the plot becoming overcrowded), which is based on the local
     dynamo simulation of \citetads{2005A&A...429..335V}. The bottom
     panel of Fig.~\ref{contint} then shows the intensity difference
     (in percentage) between our results and the observational
     dataset of \citetads{1984SoPh...90..205N}.
     The cases shown are our HD and 200 G series (the difference with
     observations being marked in the plot as black and red
     crosses, respectively). The time-averaged HD and MHD models both show a
     good fit to observations. The uncertainties involved
   in a correctly computed continuum are of the same order as the
   differences we find with respect to available literature
   constraints.

%______________________________________________________________

   \begin{figure*}[!ht]
   \centering
   \includegraphics[width=6cm]{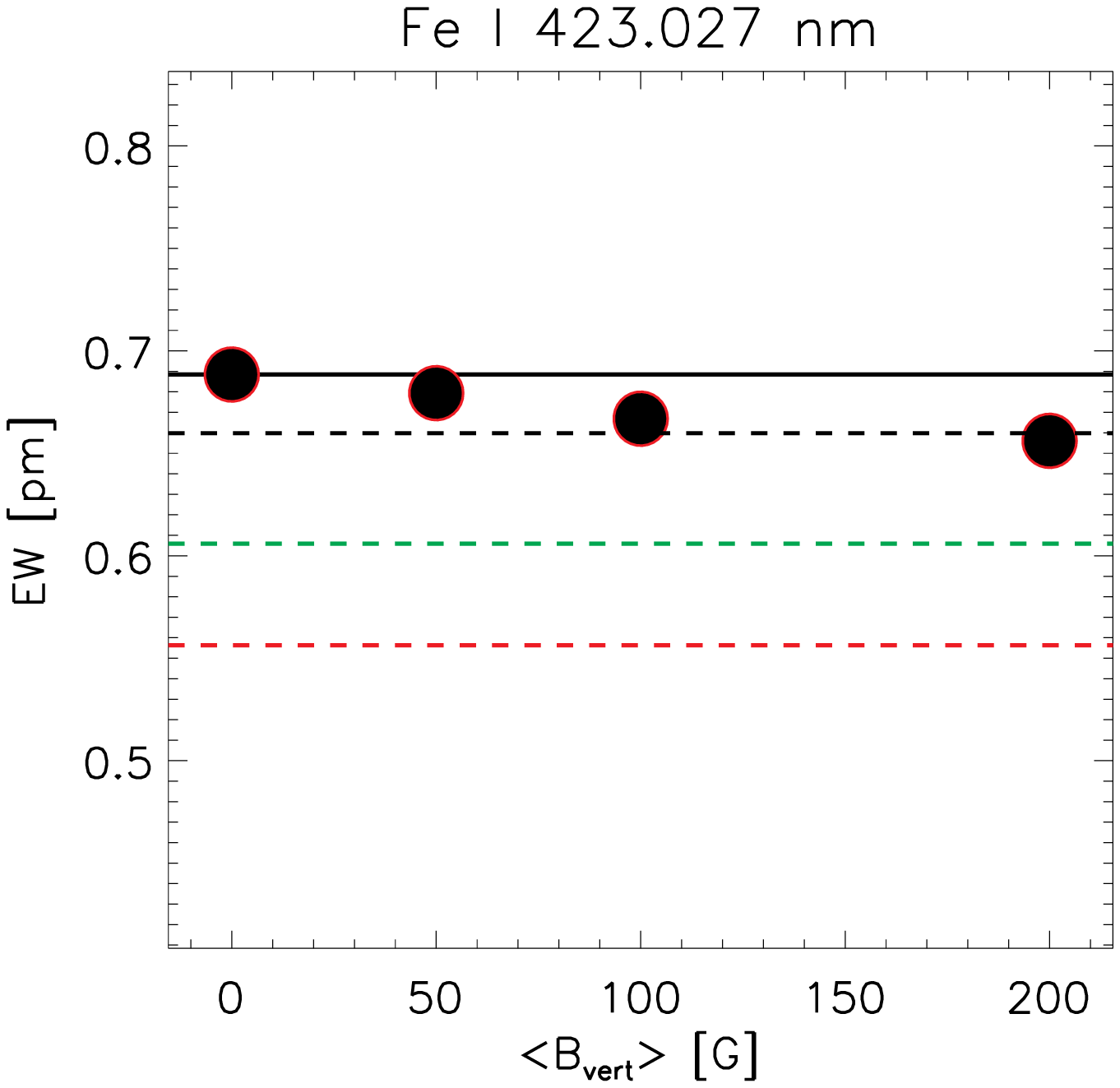} % Fe\,{\sc i} $423.027$\,nm, max $\sim -0.02$\, dex, $\mathrm{H} = ???$\,km, g_$\mathrm{L}=0$
   \includegraphics[width=6cm]{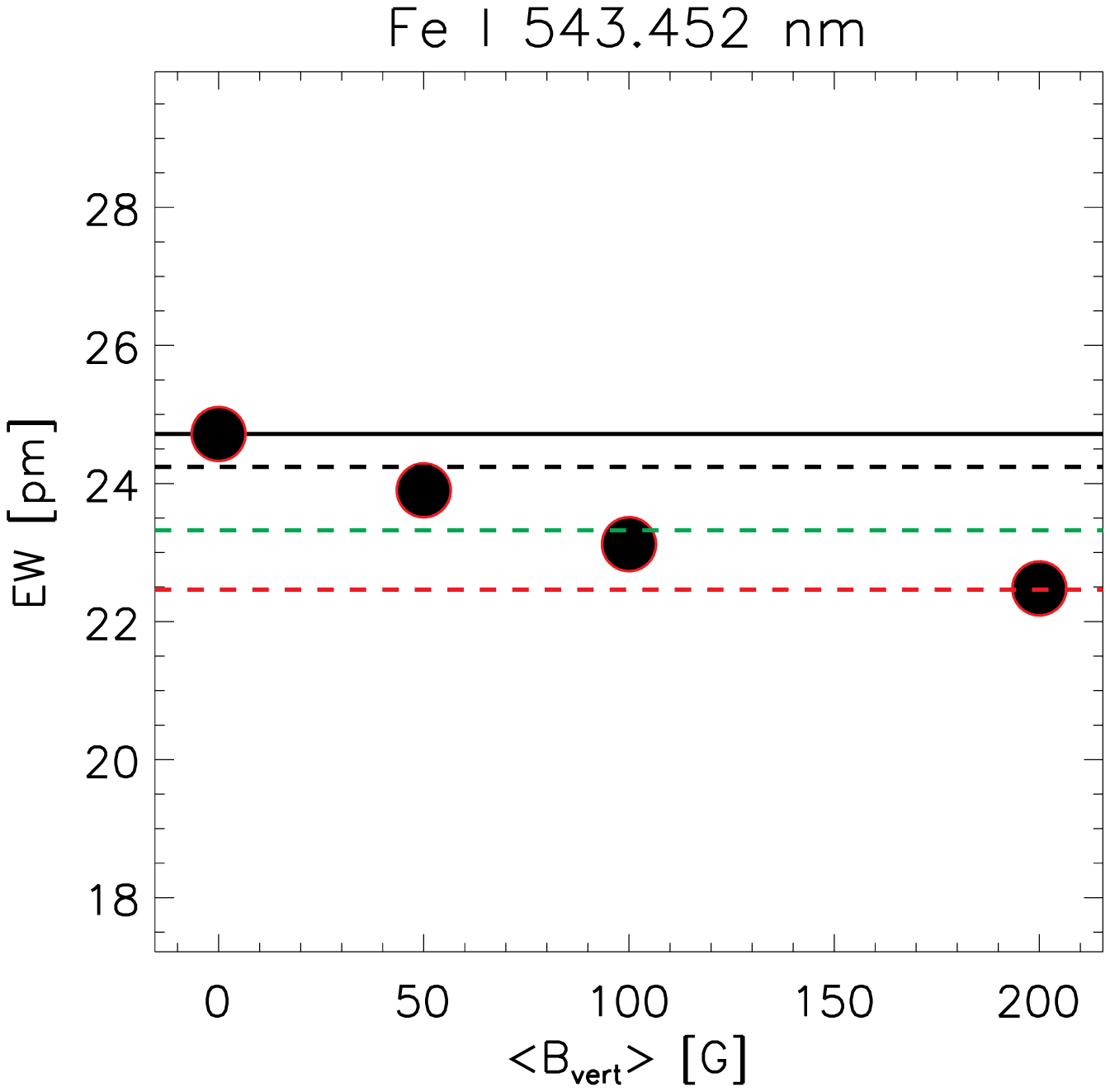} % Fe\,{\sc i} $543.452$\,nm, max $\sim -0.10$\, dex, $\mathrm{H} = ???$\,km, g_$\mathrm{L}=-0.01$ (BelloGonz)
   \includegraphics[width=6cm]{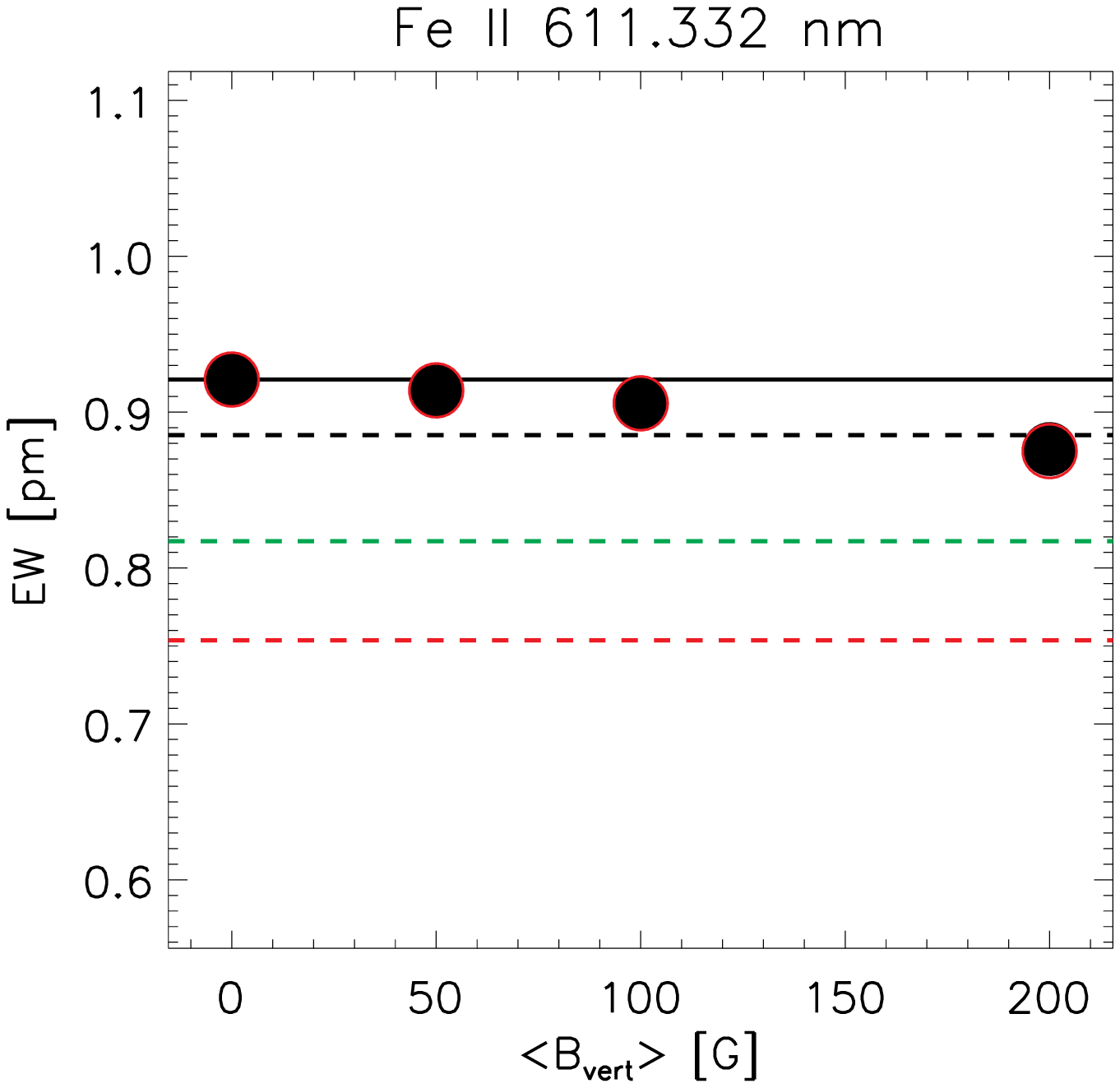} % Fe\,{\sc ii} $611.332$\,nm, max $\sim -0.03$\, dex, $\mathrm{H} = 115$\,km, g_$\mathrm{L}=0.57$
   \caption{{\it Left and middle panels:} equivalent width results for two of the Zeeman-insensitive ($g_{\mathrm{L}}=0$) spectral lines in
     our line list. {\it Right panel:} equivalent width results for the Fe\,{\sc ii}
     $611.332$\,nm spectral line.
     In both panels, filled circles mark the equivalent width for the (M)HD cases, with average flux density ($<B_{vert}> = B_{t=0} =$ const, i.~e., equal throughout the evolution to its initial value) given in abscissas. Empty circles (here, essentially overlapping the filled ones) represent the equivalent width results obtained when artificially setting $B=0$ for the spectral synthesis.
     The solid horizontal line marks the value of the equivalent width
     for the HD case. The dashed horizontal lines mark the
     equivalent width value obtained when {\it decreasing} the iron
     abundance of the HD model by $0.02, 0.06, ~\mathrm{and}~ 0.10$\, dex.}
   \label{EWs_gL_0}
   \end{figure*}

   \begin{figure*}[!ht]
   \centering
   \includegraphics[width=6cm]{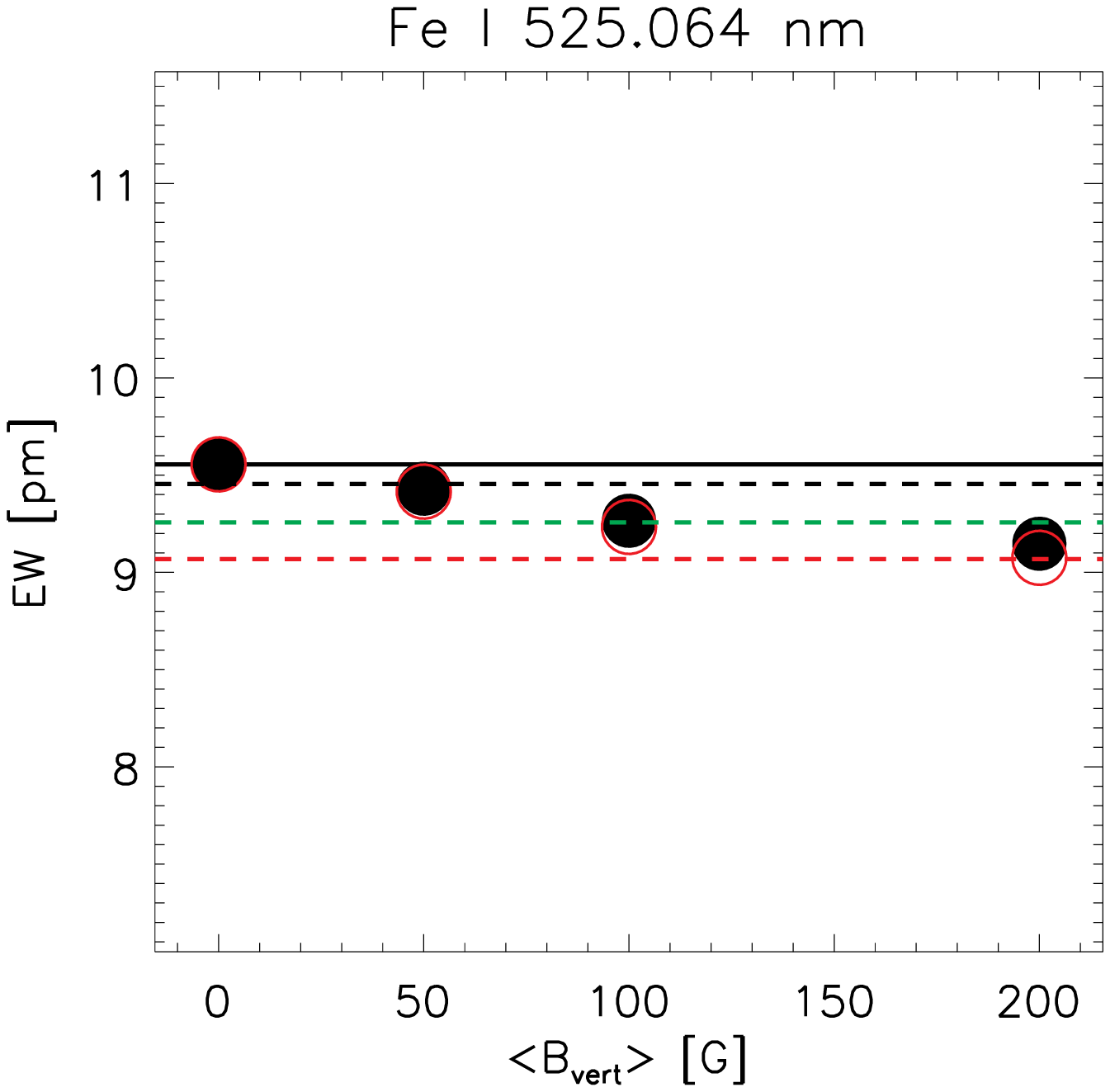} % Fe\,{\sc i} $525.064$\,nm, max $\sim -0.08$\, dex, $\mathrm{H} = 495$\,km, g_$\mathrm{L}=1.50$ % enters IMaX wavelength window (see Fig. 15 in Mart\'inez Pillet et al. (2011)) as a bonus, given that it is the previous one (525.02 nm) which was the target of the IMaX experiment.
   \includegraphics[width=6cm]{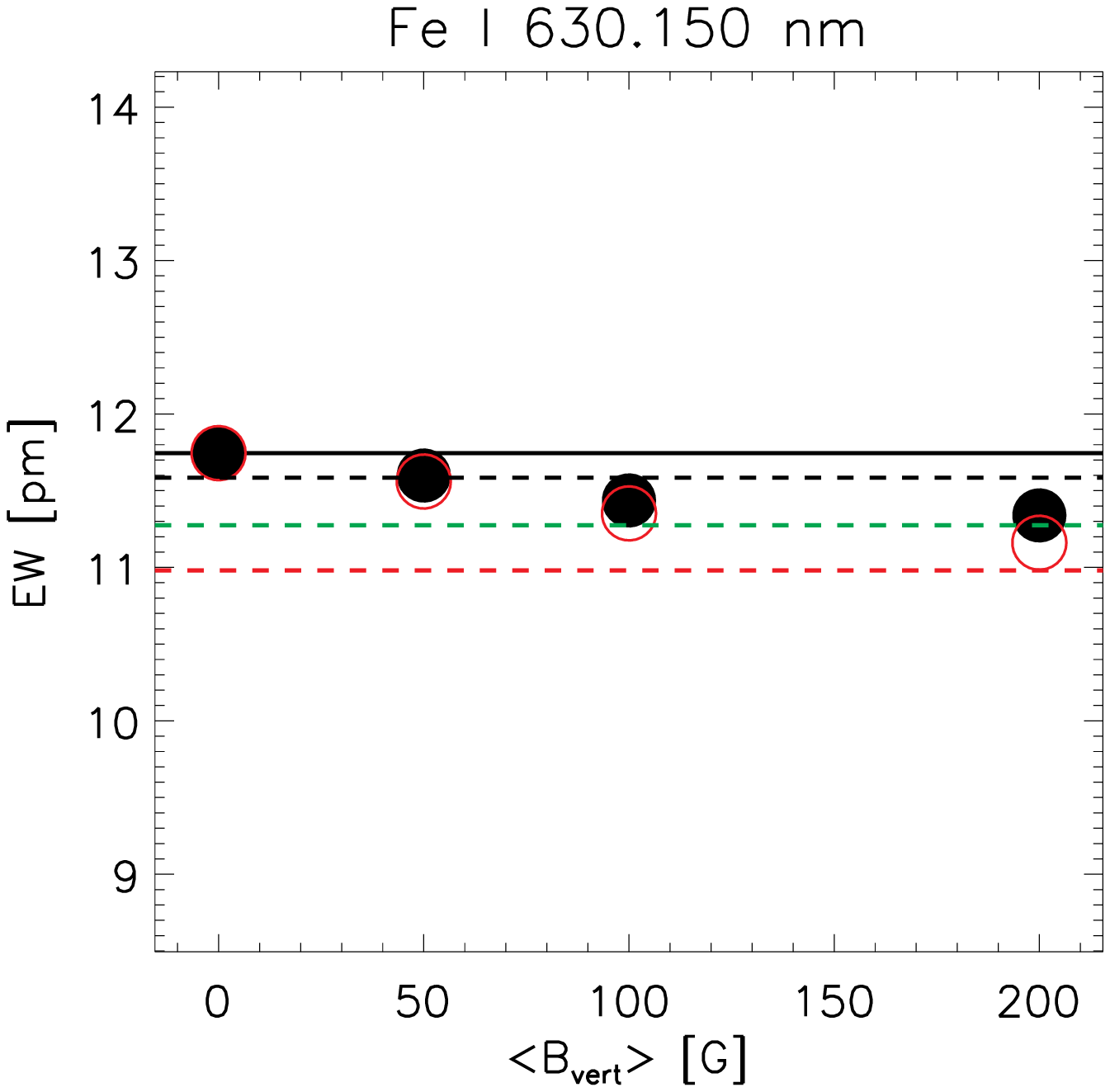} % Fe\,{\sc i} $630.150$\,nm, max $\sim -0.05$\, dex, $\mathrm{H} = 490$\,km, g_$\mathrm{L}=1.67
   \includegraphics[width=6cm]{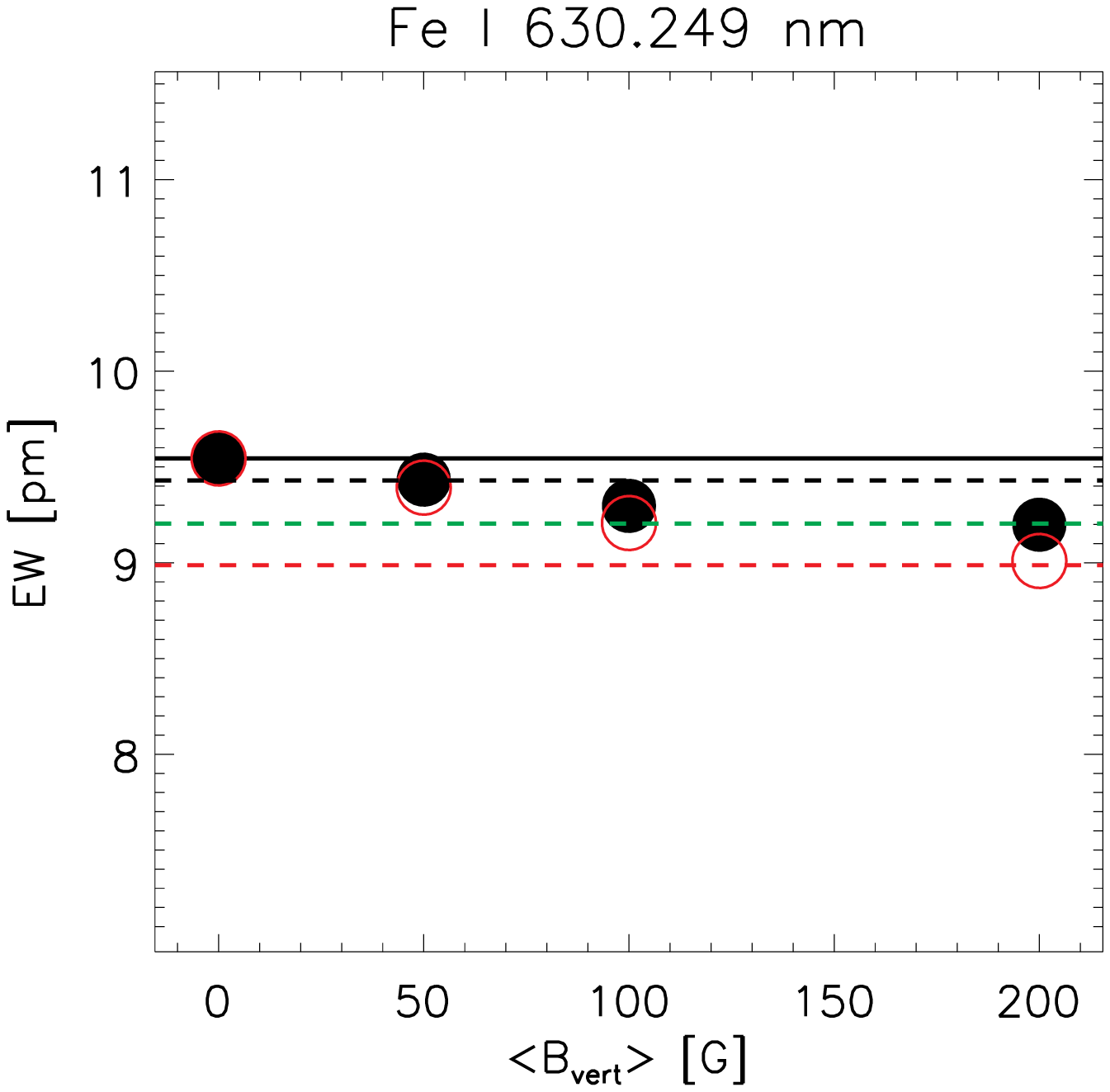} % Fe\,{\sc i} $630.249$\,nm, max $\sim -0.06$\, dex, $\mathrm{H} = 380$\,km, g_$\mathrm{L}=2.49 (e.~g., HINODE)
   \caption{Equivalent width results for three of the $g_{\mathrm{L}} > 1$} spectral features in our line
     list. The symbols are the same as in Fig.~\ref{EWs_gL_0} (but now filled and empty circles are no longer overlapping).
   \label{EWs_gL_large}
   \end{figure*}

   \begin{figure*}[!ht]
   \centering
   \includegraphics[width=6cm]{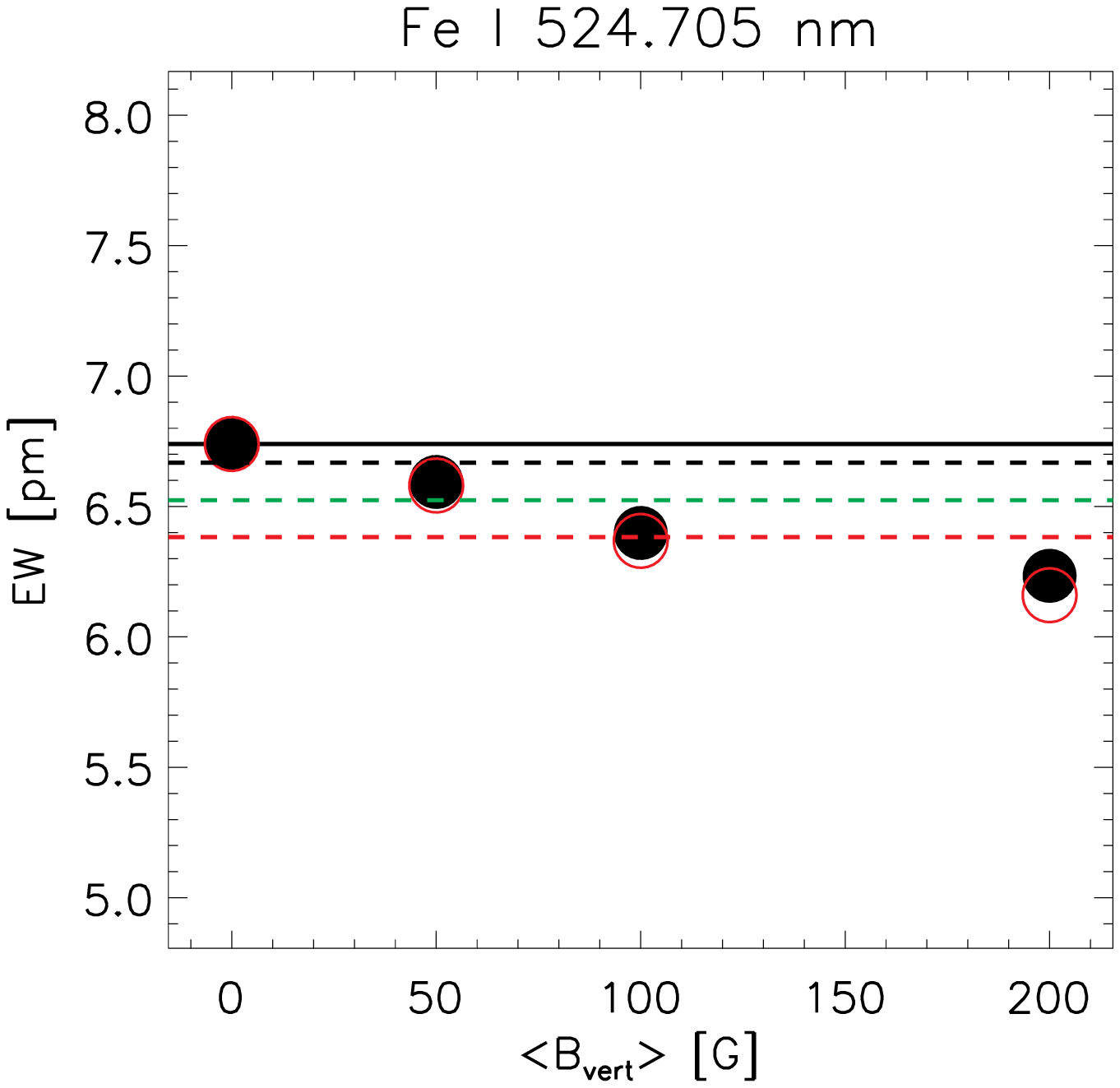} % Fe\,{\sc i} $524.705$\,nm, max $\sim -0.14$\, dex, $\mathrm{H} = 415$\,km, g_$\mathrm{L}=1.99$ (Asplund)
   \includegraphics[width=6cm]{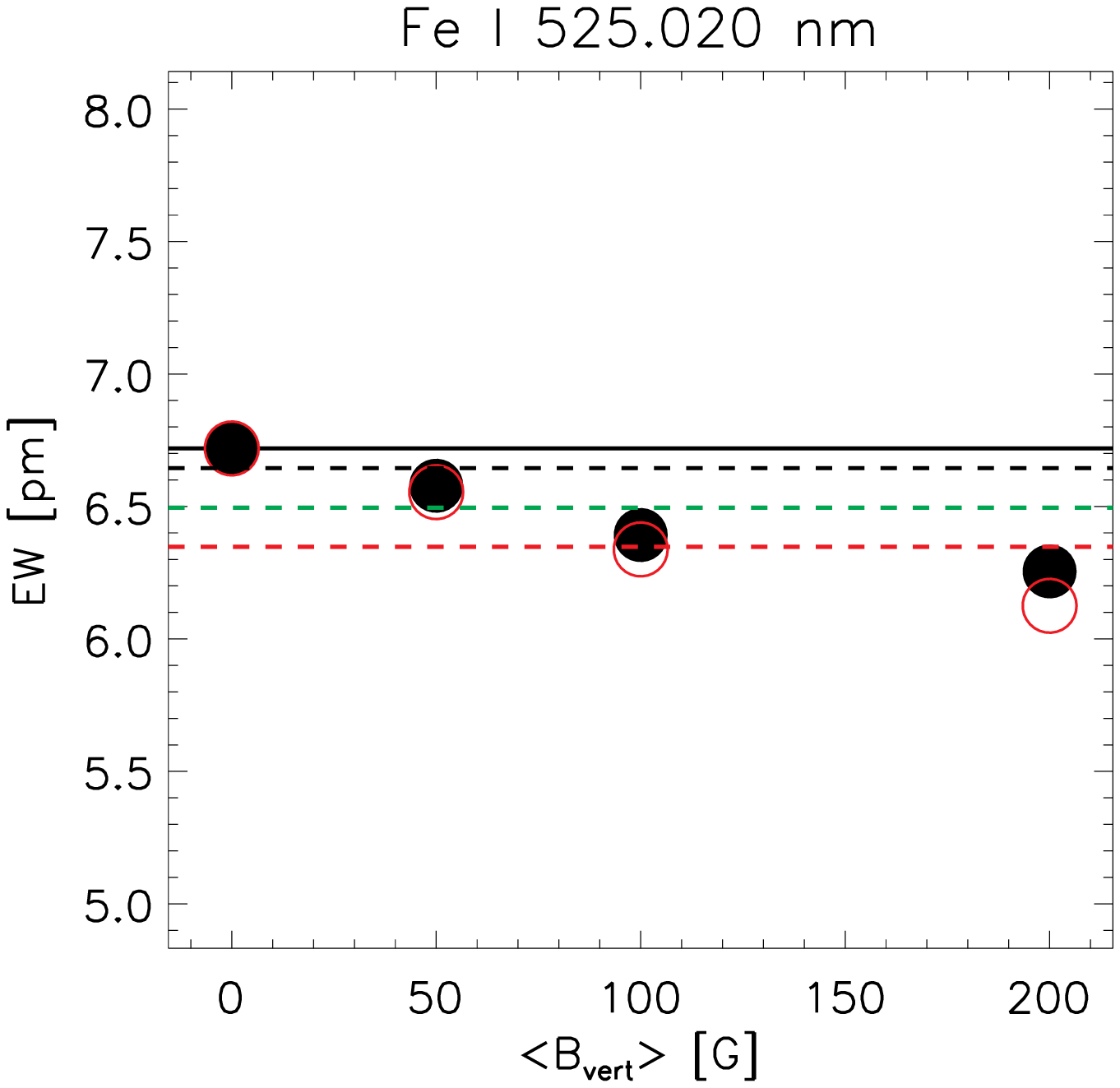} % Fe\,{\sc i} $525.021$\,nm, max $\sim -0.12$\, dex, $\mathrm{H} = 410$\,km, g_$\mathrm{L}=3.00$ (Asplund)
   \includegraphics[width=6cm]{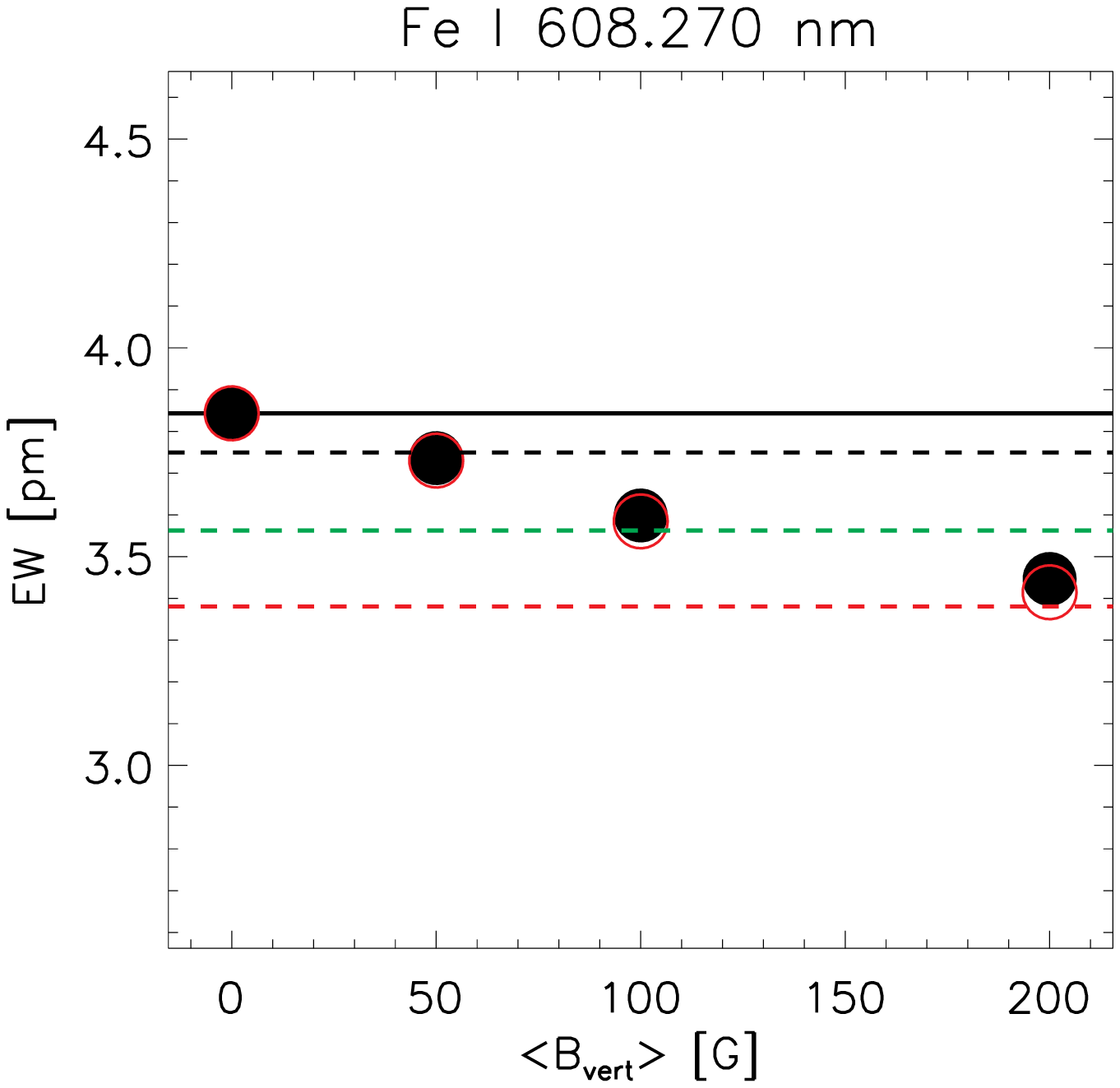} % Fe\,{\sc i} $608.271$\,nm, max $\sim -0.08$\, dex, $\mathrm{H} = 230$\,km, g_$\mathrm{L}=2.00$ (Asplund)
   \includegraphics[width=6cm]{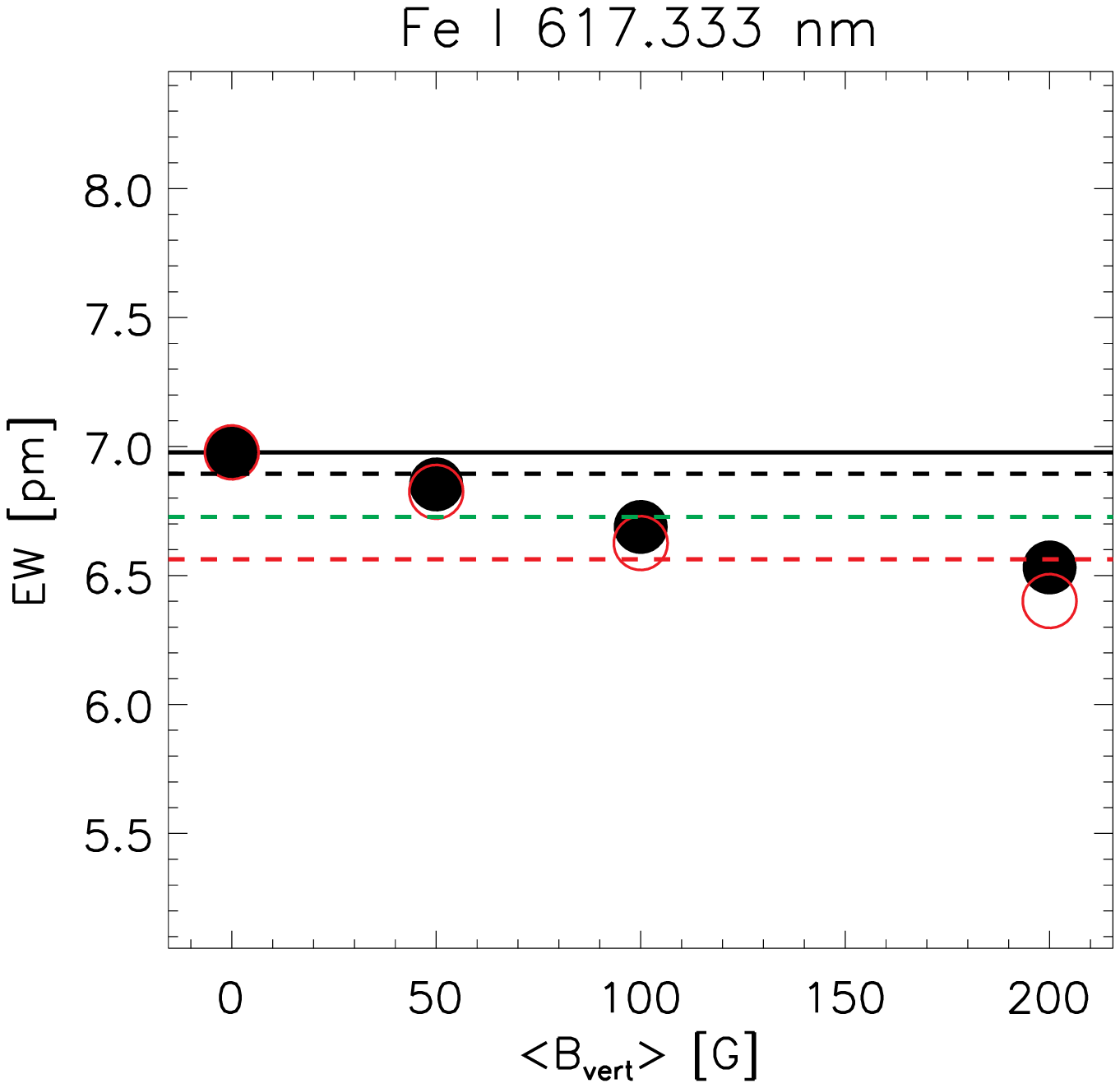} % Fe\,{\sc i} $617.334$\,nm, max $\sim -0.11$\, dex, $\mathrm{H} = 360$\,km, g_$\mathrm{L}=2.50$ (Asplund)
   \includegraphics[width=6cm]{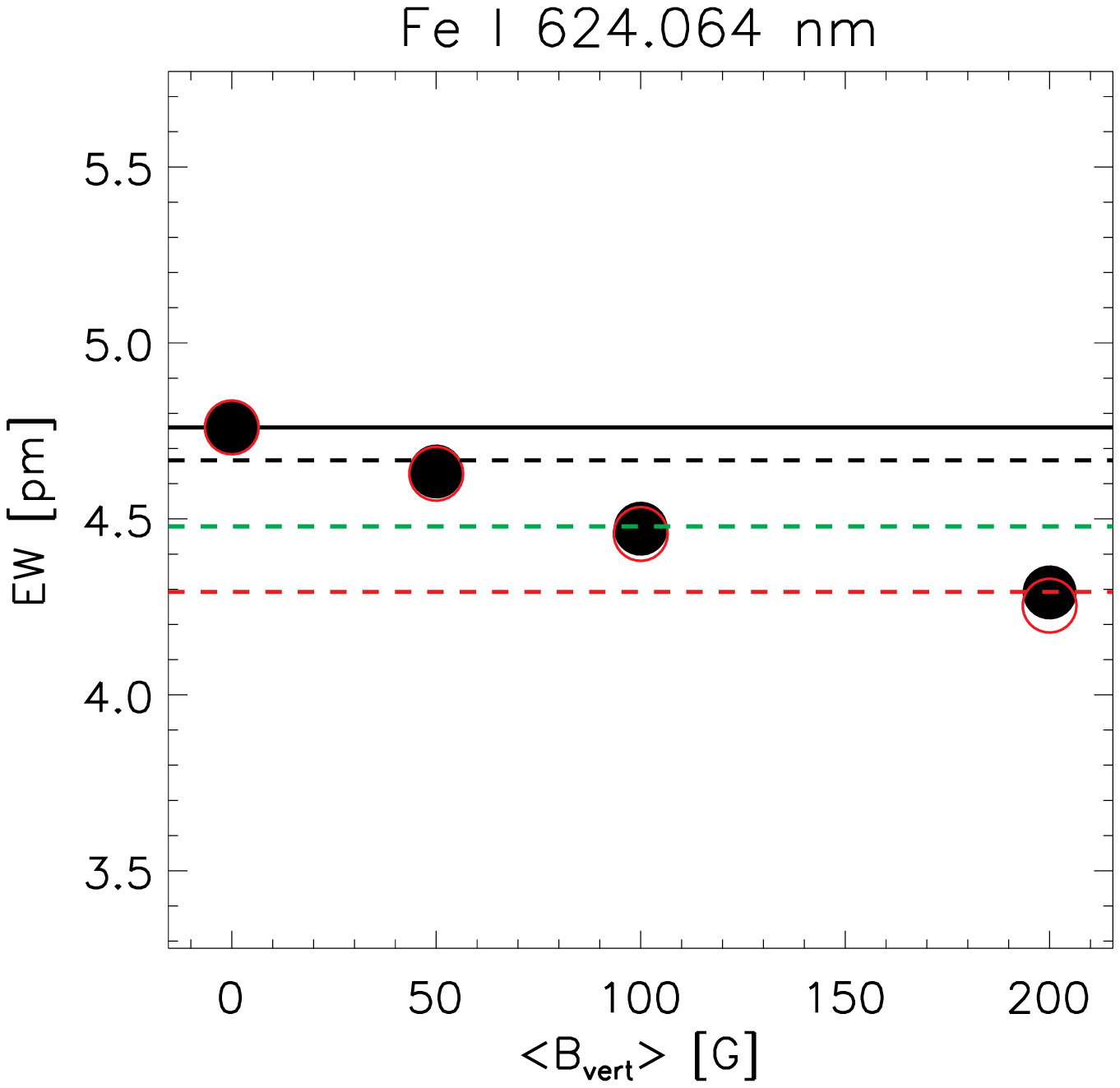} % Fe\,{\sc i} $624.065$\,nm, max $\sim -0.10$\, dex, $\mathrm{H} = 275$\,km, g_$\mathrm{L}=0.99$ (Asplund)
   \caption{Equivalent width results for the spectral lines in common
     with \citetads{2000A&A...359..743A}. Same symbols as in Fig.~\ref{EWs_gL_0}.}
   \label{EWs_Aspl}
   \end{figure*}

   \begin{figure*}
   \sidecaption
   %\centering
   \begin{minipage}{12cm}
   \resizebox{6cm}{!}{\includegraphics*{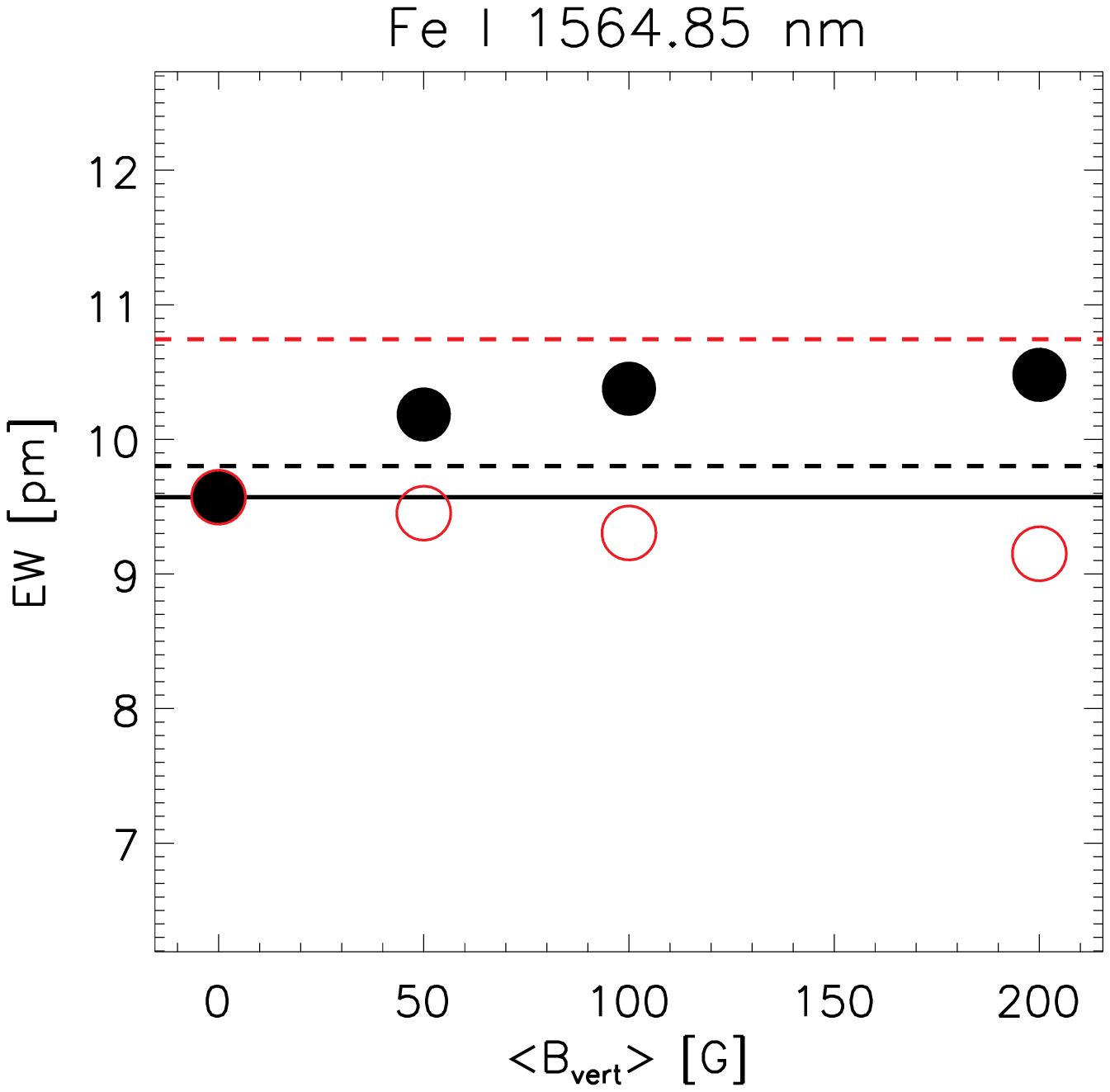}} % Fe\,{\sc i} $1564.85$\,nm, max $\sim +0.08$\, dex, $\mathrm{H} = 100$\,km, g_$\mathrm{L}=2.98$
   \resizebox{6cm}{!}{\includegraphics*{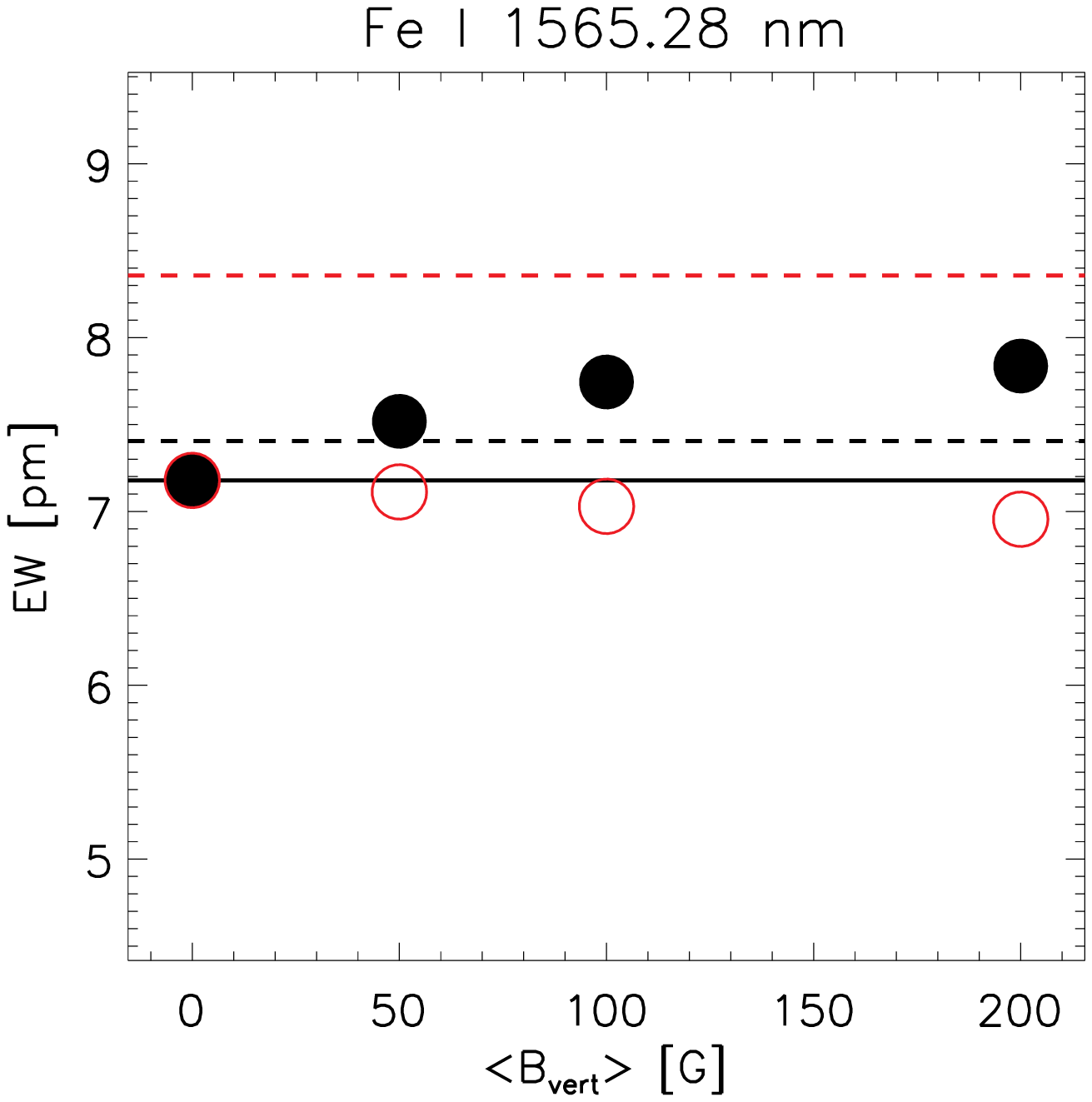}} \\ % Fe\,{\sc i} $1565.28$\,nm, max $\sim +0.06$\, dex, $\mathrm{H} = ???$\,km, g_$\mathrm{L}=1.55$
   \resizebox{6cm}{!}{\includegraphics*{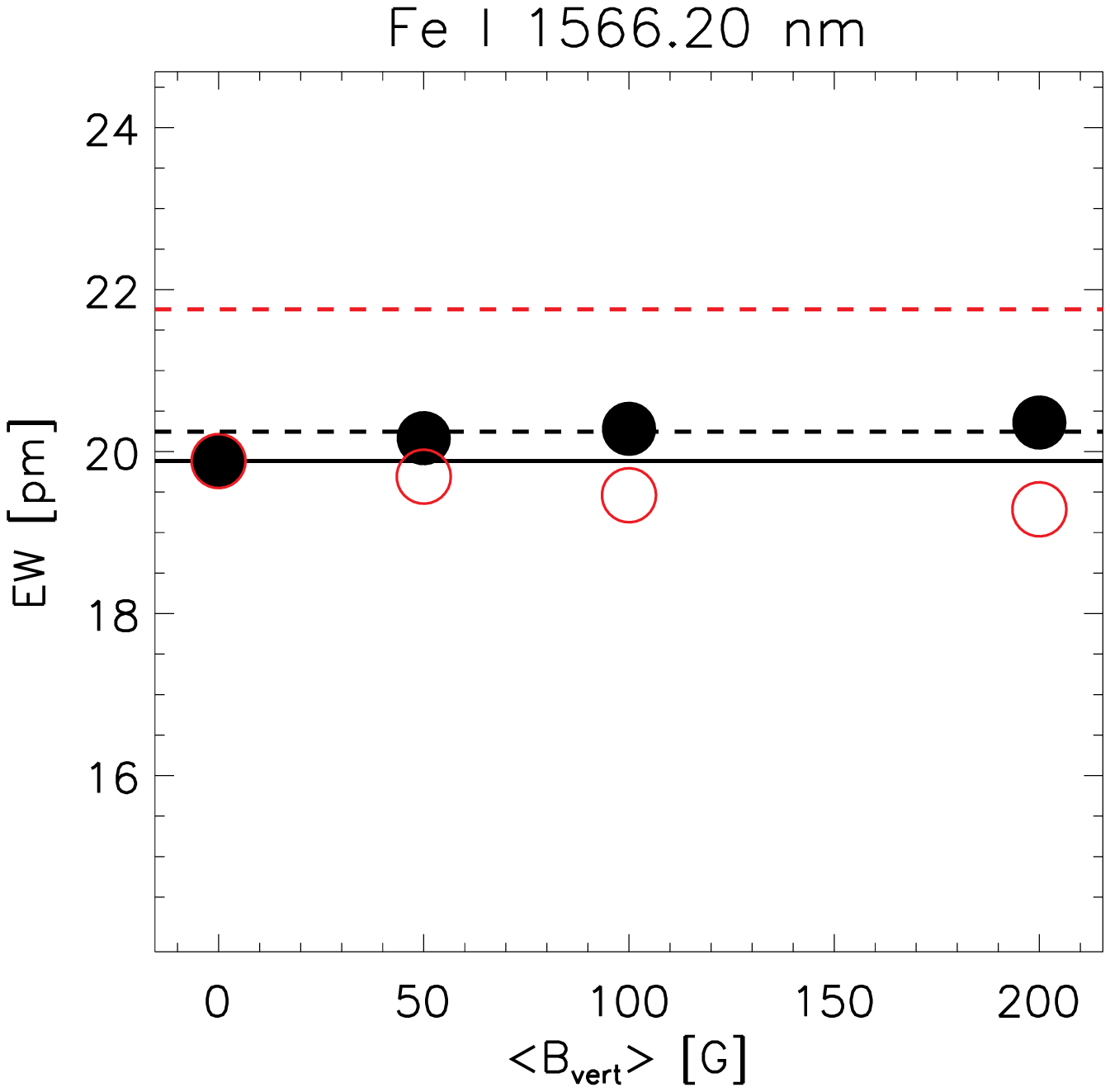}} % Fe\,{\sc i} $1566.20$\,nm, max $\sim +0.03$\, dex, $\mathrm{H} = ???$\,km, g_$\mathrm{L}=1.47$
   \resizebox{6cm}{!}{\includegraphics*{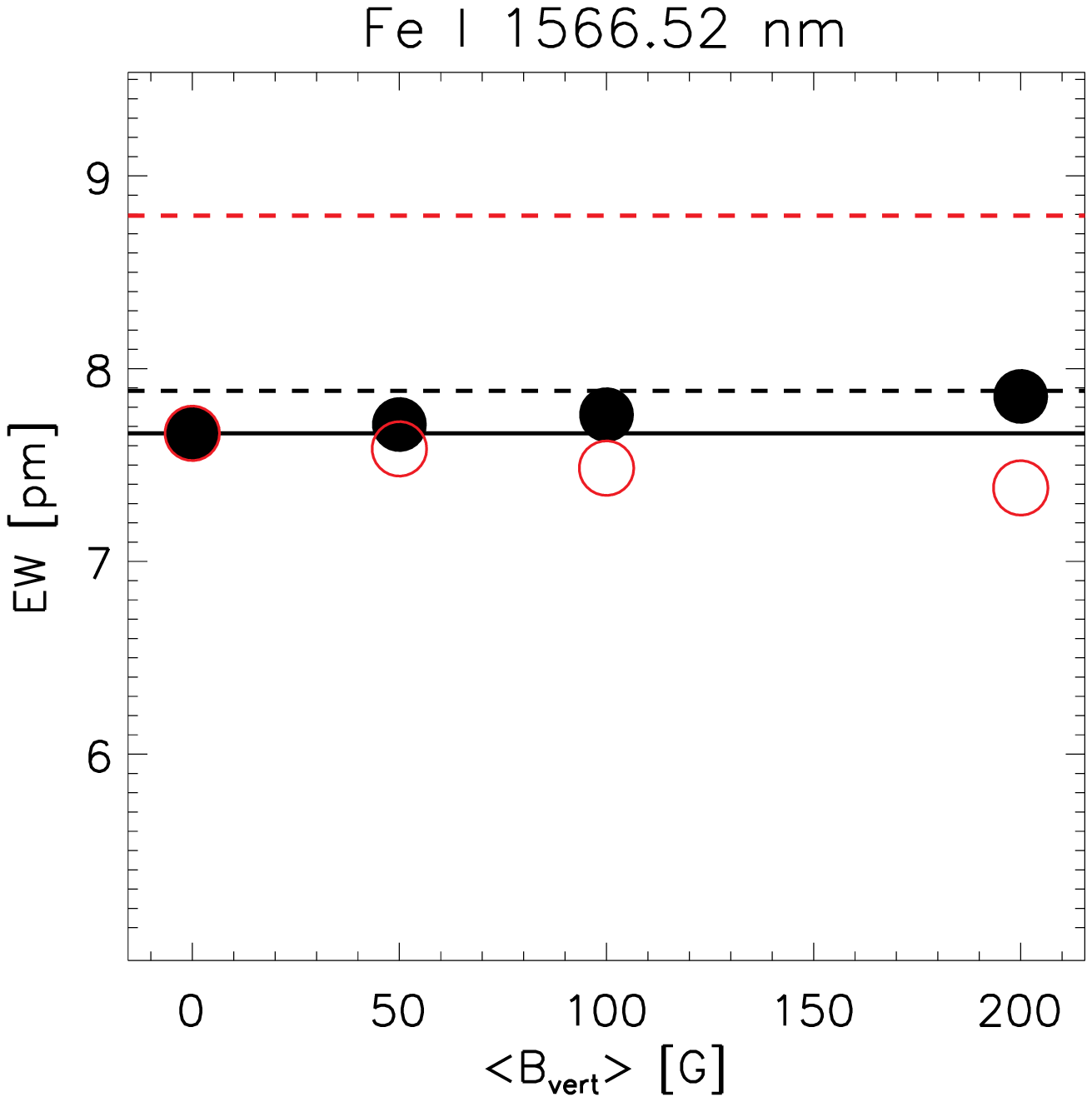}} % Fe\,{\sc i} $1566.52$\,nm, max $\sim +0.02$\, dex, $\mathrm{H} = ???$\,km, g_$\mathrm{L}=0.67$
   \end{minipage}
   \caption{Equivalent width results for the IR spectral lines
     included in our sample. The symbols are the same as in
     Fig.~\ref{EWs_gL_0}, except that now, the dashed horizontal lines
     indicate the equivalent width value predicted when {\it increasing}
     the iron abundance in the HD model by $0.02 ~\mathrm{and}~ 0.10$\, dex. This translates to {\it negative} abundance corrections of
     the same magnitude applicable to the abundance predicted using
     the HD model (see discussion in text).}
   \label{EWs_IR}
   \end{figure*}

\section{Abundance correction results}
\label{section:abucorr}

\begin{table*}
  \caption{Abundance corrections for the Fe spectral lines in the visible
    range, for the different magnetic flux cases.}
\label{abucorr_visible}      % is used to refer this table in the text
\centering                          % used for centering table
\begin{tabular}{c c c c c c}        % centered columns (6 columns)
\hline\hline                 % inserts double horizontal lines
$\lambda$ & $\Delta \mathrm{\log} \, \epsilon \mathrm{(Fe)}_{\odot}$ & $\Delta \mathrm{\log} \, \epsilon \mathrm{(Fe)}_{\odot}$ & $\Delta \mathrm{\log} \, \epsilon \mathrm{(Fe)}_{\odot}$ & W$_{\mathrm{HD}}$ & ID  \\    % table heading
\protect  [nm]  & [dex]     &   [dex]     &   [dex]    & [pm] &   \\
                & ($50$\,G - HD) &  ($100$\,G - HD) & ($200$\,G - HD) &  &   \\
\hline                        % inserts single horizontal line
  410.091                &  $+0.01$ & $+0.03$  &    $+0.04$  &  0.55  & 0L \\
  415.025                &  $+0.02$ & $+0.04$  &    $+0.06$  &  7.94  & 0L \\
  423.027                &  $+0.01$ & $+0.01$  &    $+0.02$  &  0.69  & 0L \\
  430.603                &  $+0.01$ & $+0.02$  &    $+0.03$  &  1.98  & 0L \\
  437.113                &  $+0.01$ & $+0.01$  &    $+0.02$  &  0.10  & 0L \\
  486.364                &  $+0.02$ & $+0.05$  &    $+0.09$  &  6.60  & 0L \\
  524.705 & $+0.04$ & $+0.09$ & $+0.14$       &  6.74  & A \\
  525.021 & $+0.04$ & $+0.08$ & $+0.12$       &  6.72  & A \\
  525.065                & $+0.03$ & $+0.06$  &    $+0.08$   &  9.56  & LL \\
  543.452                & $+0.04$ & $+0.07$  &    $+0.10$   & 24.72  & 0L \\
  557.609                & $+0.02$ & $+0.05$  &    $+0.07$   & 10.91  & 0L \\
  579.119                & $+0.01$ & $+0.02$  &    $+0.03$   &  0.37  & 0L \\
  585.923                & $+0.01$ & $+0.02$  &    $+0.03$   &  0.18  & 0L \\
  594.036                & $+0.01$ & $+0.02$  &    $+0.03$   &  0.25  & 0L \\
  601.392                & $+0.01$ & $+0.02$  &    $+0.03$   &  0.28  & 0L \\
  608.271 & $+0.02$ & $+0.05$ & $+0.08$       &  3.84  & A \\
  611.332 & $ 0.00$ & $+0.01$ & $+0.03$       &  0.92 & II \\
  616.946                & $+0.01$ & $+0.03$  &    $+0.05$   &  0.16  & 0L \\
  617.334 & $+0.03$ & $+0.07$ & $+0.11$       &  6.98  & A \\
  624.065 & $+0.03$ & $+0.06$ & $+0.10$       &  4.76  & A \\
  630.150                & $+0.02$ & $+0.04$  &    $+0.05$   & 11.75  & LL \\
  630.249                & $+0.02$ & $+0.04$  &    $+0.06$   &  9.55  & LL \\
  633.084                & $+0.01$ & $+0.02$  &    $+0.04$   &  1.37  & LL \\
\hline
  $\langle \rangle$     & $0.019    $ &   $0.040    $ & $0.061    $ &   &   \\
                        & $\pm 0.011$ &   $\pm 0.023$ & $\pm 0.034$ &   &   \\
\hline
  $\langle \rangle$*    & $0.032    $ &   $0.070    $ & $0.110    $ &   &   \\
                        & $\pm 0.008$ &   $\pm 0.014$ & $\pm 0.020$ &   &   \\
\hline\hline
\end{tabular}
\tablefoot{Columns in the top part list the rest wavelength of the absorption lines and their abundance corrections for the three magnetic flux cases, the equivalent width obtained from the spectral
    synthesis based on our HD series, and the identifiers associated with each spectral line,
    following the convention introduced in Table~\ref{lines_params}. In the bottom part, $\langle \rangle$ indicates the unweighted mean of the abundance corrections for the visible
    lines and corresponding {\it rms} for the different magnetic flux cases, and $\langle \rangle$* refers to the values obtained when averaging for just
    the 5 lines in common with \citetads{2000A&A...359..743A}.\\
}
\end{table*}

\subsection{Overview}
\label{ss:over}

The procedure used to obtain abundance corrections is as follows.
From the (M)HD snapshots we obtained synthetic spectra for each of the
columns in the snapshots and for all of the lines in Table~\ref{lines_params}
using a standard reference abundance value of $\mathrm{\log} \, \epsilon
\mathrm{(Fe)}_{\odot} = 7.50$\, dex (although the precise value is not crucial
here, given the differential nature of our study). Then, we calculated space
and time averages (i.e., horizontally in each snapshot and for all snapshots)
separately for each of the series (HD, $50$ G, $100$ G, $200$ G). We then
computed the equivalent width of the average profiles and compared them
across the series.  Additionally, to determine abundance corrections, we ran
spectral synthesis calculations for the HD case with an input Fe
abundance changed by $-0.10, -0.06, -0.02, +0.02,~{\mathrm{and}}~+0.10$\, dex compared
to the reference value mentioned above.
The precise abundance correction to use in the HD series to match the equivalent
width of the profiles for the MHD series was then found by interpolation
between the values obtained for these fixed abundance jumps. In fact, the
abundance correction to adopt as a result is actually the negative of the
value obtained through the interpolation.
The reason for the change of sign is that if,
e.g., a given spectral line tends to weaken for non-zero magnetic flux,
then when neglecting magnetic fields the equivalent width measured from observations
will be matched with a lower (i.~e.  {\it underestimated}) elemental
abundance.

Finally, to separate the magnetic broadening (Zeeman splitting)
effects from the influence of the changed temperature structure in the
MHD models, we ran further {\texttt{LILIA}} spectral synthesis
calculations based on the MHD models (in particular, keeping the temperature
and density stratification in them) but artificially setting $B=0$.
We expect that the line formation in this case will contain the indirect
effect of the magnetically-induced temperature stratification change, but of
course, without any direct Zeeman broadening.

In the following sections, we present the results of this study by
dividing the spectral features in our line list into two broad
groups. The first group (Sect.~\ref{sec:gL_eq0})
contains the 13 lines with no
Zeeman sensitivity (g$_\mathrm{L} \sim 0$), while the second group
(Sect.~\ref{sec:gL_neq0}) is
composed of the 15 lines with at least some direct sensitivity ($0.5
< g_\mathrm{L} \le 3.0$) to the magnetic field.

\subsection{Zeeman-insensitive spectral lines ($g_L=0$)}
\label{sec:gL_eq0}

Spectral lines with $g_L=0$ are indicated with ``0L'' in
Tables~\ref{lines_params} and \ref{abucorr_visible}. Any magnetic
effects on these lines must necessarily come from the change of
temperature and density stratification alone. The magnitude of the
effect then depends on the formation height and temperature
sensitivity of the line. The behaviour of the calculated equivalent
width with respect to magnetic flux is illustrated for two lines of
this group in Fig.~\ref{EWs_gL_0} (left and middle panels).  The
corresponding abundance corrections are included in
Table~\ref{abucorr_visible}, together with the predicted HD equivalent
width for each spectral line (which can provide an idea of their 
approximate formation region, since the formation height is not available 
from \citeads{1989GurtKost} for all lines in our list). 
The abundance corrections turn out to be
small ($\lessapprox +0.02$\, dex) for \mbox{Fe\,\sc{i}} $423.027$\,nm
(left panel), but reach up to $\sim +0.10$\, dex for \mbox{Fe\,\sc{i}}
$543.452$\,nm (middle panel) in the 200 G case.  The hotter
temperatures that the radiation ``feels'' in the MHD models make
(for a fixed adopted abundance) these spectral lines increasingly weaker the higher the unsigned flux of the different MHD
series, in agreement with their temperature sensitivity.  By checking
the position of the dashed horizontal lines, we see that the abundance
of the HD case must be {\it decreased appreciably} for it
to match the equivalent width found for the MHD cases. Using the non-magnetic case to match measured equivalent widths from
  observed spectra, one would then derive a low abundance estimate.
  The latter will in fact allow to match predicted and measured equivalent
  widths because the spectral line is intrinsically stronger in
  HD. This finally translates into a {\it positive} abundance
  correction to the purely HD case, because the abundance values
  obtained for it must be revised upward in order to match
  observations while also endeavouring to account (precisely via the
  use of the relevant abundance correction) for the missing effects
  (instead naturally present in MHD) due to magnetic fields.

   The other Zeeman-insensitive spectral lines in our sample all have
   resulting abundance corrections within the above values. For
   example, they reach up to only $\sim +0.03$\, dex for the
   Fe\,{\sc i} $585.923$\,nm feature, but up to $\sim +0.09$\, dex for the
   $486.364$\,nm line.

   As seen in Fig.~\ref{EWs_gL_0} (left and middle panels), 
   the Zeeman-insensitivity of all
   absorption lines in this category is confirmed by the absence of
   any equivalent width variation when artificially setting $B=0$ in
   the spectral synthesis.

\subsection{Spectral lines with g$_{\mathrm{L}} \neq 0$}
\label{sec:gL_neq0}

\subsubsection{Fe\,{\sc ii} $611.3$\,nm  line}

Figure~\ref{EWs_gL_0} (right panel) shows the equivalent width
results for the only Fe\,{\sc ii} spectral line included in this study
(marked with ``II'' in Tables~\ref{lines_params} and
\ref{abucorr_visible}). The corresponding corrections to the
HD-derived abundance are modest.  Since Fe\,{\sc ii} lines are weak
(our predicted HD equivalent width for this line is $0.92$~pm, as listed in Table~\ref{abucorr_visible}) and generally formed rather deep in the photosphere, in the region
where the temperature modifications due to the presence of magnetic
fields are small (see
Fig.~\ref{fig1_avTxyt_vsDepth_and_vsTau_ks1in5}), the corresponding
abundance corrections can also be expected to be fairly mild.  The
formation height of the Fe\,{\sc ii} $611.332$\,nm line lies around
$110$\,km (\citeads[see Table 2 in ][]{1989GurtKost}).  This line, 
moreover, belongs to the visible wavelength range and has a small
Land\'e factor (g$_\mathrm{L}=0.57$), so that, as seen in
  Fig.~\ref{EWs_gL_0} (right panel), the Zeeman broadening is unable
to contribute any significant line strengthening. The applicable
abundance corrections thus settle at a maximum of $\sim +0.03$\, dex,
see Table~\ref{abucorr_visible}. Singly-ionized iron lines are
generally less sensitive to temperature, to non--LTE effects and to
details of the model atmosphere \citepads[see,
e.~g.,][]{2009ARA&A..47..481A}. Improved atomic data for them are now
available (e.g., \citeads{2009A&A...497..611M}). It will therefore be of future
interest to ascertain, based on a larger sample of singly-ionized iron
lines, whether they may also be generally insensitive to the direct
and indirect effects of magnetic fields, as the
results for at least Fe\,{\sc ii} $611.332$\,nm seem to suggest. This
would likely confirm Fe\,{\sc ii} as a more reliable indicator than
Fe\,{\sc i} for the purpose of abundance determinations under
simplified (e.g., LTE, or non-magnetic) approximations, i.e., when a
more sophisticated analysis is not feasible.

\subsubsection{Lines with large Land\'e factor $g_{\mathrm{L}}$}
\label{sec:lines_large_lande}

Spectral lines with $g_L \ge 1$ are indicated with ``LL'' in
Tables~\ref{lines_params} and ~\ref{abucorr_visible}.  The abundance
corrections for these lines are given in Table~\ref{abucorr_visible},
together with their predicted HD equivalent width.
Figure~\ref{EWs_gL_large} shows the equivalent width results for three
spectral features with large $g_{\mathrm{L}}$, namely the
$630.150$\,nm, $630.249$\,nm, and $525.064$\,nm lines.  The abundance
corrections for these lines are of intermediate magnitude, reaching
the value $\sim +0.05$\, dex, $\sim +0.06$\, dex, and $\sim +0.08$\, dex,
respectively. Given their non-zero Zeeman sensitivity, the equivalent
width of all three lines decreases when artificially setting $B=0$
for the spectral synthesis, as seen in the figure. However, this effect remains small throughout
the visible range, and is not sufficient to counter the indirect
line-weakening temperature effect.

   Some of the lines marked as A or IR in
   Table~\ref{lines_params} also have a large Land\'e
   factor $g_{\mathrm{L}}$. However, they are discussed as separate
   subgroups below, because they are helpful for highlighting some of the
   main results of this study, i.e. the possible
   impact on the accepted value of the solar iron abundance
   and the more noticeable Zeeman broadening effects in
   the IR, respectively.

\subsubsection{Lines in common with the solar iron abundance study of Asplund et al. (2000)}\label{sec:lines_common_with_asplund}

Figure~\ref{EWs_Aspl} shows the equivalent width results for the
spectral features marked with ``A'' in Tables~\ref{lines_params} and
\ref{abucorr_visible}. We include them in an own separate subgroup,
given that they are the ones in common with a previous well-known
solar Fe abundance study \citepads{2000A&A...359..743A}. Corresponding
abundance corrections and HD equivalent widths are listed in
Table~\ref{abucorr_visible}.  These spectral lines have some of the
largest Land\'e factors among the visible lines in our sample. 
However, since all five are in the visible range, Zeeman broadening
actually has a little influence on them and can do only little to
alleviate the line weakening tendency, as is apparent by comparing the
empty and filled circles in the panels.  On the other hand, their
sensitivity to the magnetically induced changes in the average
temperature stratifications is high, placing these spectral features
among those in our line list that have the largest abundance corrections
(see Table~\ref{abucorr_visible}). Namely, for the 200 G case,
  the corresponding correction to the HD-derived abundance is between
$0.08$\, dex for the $608.271$\,nm line to $0.14$\, dex for the
$524.705$\,nm one, which is the maximum in the whole sample we
studied.

Fe\,{\sc i} $525.021$\,nm (one of the two spectral lines in the ``good
magnetic pair'' of \citeads{Stenflo1973}) has the second most
significant corrections, $\Delta \mathrm{\log} \, \epsilon
\mathrm{(Fe)}_{\odot}=0.12$\, dex.  Apart from having been used in the
solar iron abundance derivation of \citetads{2000A&A...359..743A},
this line has also recently been targeted by the IMaX imaging vector
magnetograph instrument (\citeads{2011SoPh..268...57M};
\citeads{2010ApJ...723L.139B}) onboard the ballon-borne Sunrise
telescope mission (\citeads{2011SoPh..268....1B};
\citeads{2010ApJ...723L.127S}).

\subsubsection{IR lines}\label{sec:IR_lines}

   Spectral iron lines belonging to the infrared wavelength range
     are identified with ``IR'' in Table~\ref{lines_params}.
   Table~\ref{abucorr_IR} lists their abundance corrections and HD
   equivalent widths. All of the IR lines included in our line
     list are strong, $7.6 < {\rm W}_{\rm HD} < 19.9$~pm.
   Figure~\ref{EWs_IR} shows the relevant equivalent width results.
   As seen, these lines behave like those in the visible concerning
   their sensitivity to the warmer temperature structure in the MHD
   models. However, the IR iron lines we selected have quite
     large Land\'e factors. Coupled with the $\lambda^{2}$ dependence
     of Zeeman broadening, this makes them prone to strong direct
     magnetic effects. The result is that, compared to the HD
     case, these IR iron lines tend to generally become stronger in
     the MHD models. The corresponding abundance corrections range
   from negligible to $-0.08$\, dex.

\section{Consequences concerning the solar Fe abundance}
\label{section:consequences}

Different neutral and/or ionized iron lines should yield a consistent
estimate for the photospheric solar abundance. Once using
appropriately weigthed averaging of the abundance values inferred from
the various spectral lines of a given chemical element, little scatter
should in fact result around the ``best estimate''. In this section,
we provide average values for the abundance correction to be applied
to different groups of lines. In doing so, we just conform to the
usual practice in the literature whereby averages of abundance values
from multiple spectral lines have been routinely taken as the
representative derived solar abundance. Since there is no obvious way
to attach a different statistical significance to the abundance
correction obtained for each line, we simply give equal weight to them
when calculating averages.\\

{\bf Spectral lines in the visible}: as can be seen in
Table~\ref{abucorr_visible}, the abundance corrections for the lines
in the visible range vary from negligible to as high as $+0.14$\, dex.
The unweighted mean of the corrections for these lines is shown in the
row near the bottom marked with $\langle \rangle$ signs and it varies
between $0.02$ and $0.06$\, dex when going from the 50-G to the 200-G
case.  If we focus on the visible Fe lines marked with ``A'' in our
table (i.e., those in common with \citetads{2000A&A...359..743A}), we
find that they yield significantly higher abundance values when MHD
modelling is used. In detail, out of these five spectral lines,
Fe\,{\sc i}\,$608.271$\,nm is the least affected, with abundance
corrections of $0.02-0.08$\, dex, depending on the magnetic flux
considered. The most affected is Fe\,{\sc i}\,$524.705$\,nm, with
abundance corrections of $0.04-0.14$\, dex. 

We may apply the average abundance correction values (given in the row
marked with $\langle\rangle*$ in Table~\ref{abucorr_visible}) to the
abundance determinations by \citetads{2000A&A...359..743A} for those
five lines. Their average value was $\mathrm{\log} \, \epsilon
\mathrm{(Fe)}_{\odot} = 7.42 \pm 0.03$. We would then obtain an
average solar iron abundance estimate in the range $7.45 \le
\mathrm{\log} \, \epsilon \mathrm{(Fe)}_{\odot} \le 7.53$\, dex,
depending on the magnetic flux assumed. On the other hand, even if we
apply abundance corrections separately for the different lines
following the results in Table~\ref{abucorr_visible} and calculate the
mean value and {\it rms}, the value of the latter turns out to be close to the one
calculated using purely HD models. In detail: the {\it rms} is
$0.03$\,dex for the 50 G case and $0.04$\,dex for the 100 and 200 G
cases, hence very similar to the values obtained by
\citetads{2000A&A...359..743A}.  Therefore, the good line-to-line
agreement found by those
authors is not spoilt when the MHD approach is considered.\\

{\bf Spectral lines in the IR}: the average value of the abundance
corrections for the IR iron lines we studied is indicated in
Table~\ref{abucorr_IR} (bottom row, marked with $\langle\rangle$). We
derive a {\it negative} average correction in the range $-0.02 <
\langle\Delta \mathrm{\log} \, \epsilon \mathrm{(Fe)}_{\odot}\rangle <
-0.04$\,dex. However, the scatter around the mean is quite large for
these lines, because of the scatter in their Land\'e factor
(see Table~\ref{lines_params}) and the relative importance that the
Zeeman effect has for each of these IR lines.

\section{Discussion}
\label{section:discussion}

First results using a 3D MHD approach to abundance analysis were
presented in our previous paper \citep{2010ApJ...724.1536F}, where the
study was limited to three iron lines. Here, we have aimed at extending
those exploratory results using a much larger set of spectral lines,
which allowed us to further improve the constraints on the relevant
derived effects on the solar abundance. However, it is important to
note that, given the differential nature of our study, it is not our
main aim to match the exact value of measured equivalent widths.
Despite this, with the input solar iron abundance we adopted for
  the spectral synthesis and using a sensible choice of magnetic flux,
  we find generally good agreement between our predicted MHD
  equivalent widths and those from observational data available in the
  literature for lines marked ``A'' in our Table~\ref{abucorr_visible}
  (e.g., the value by \citeads{1995A&A...296..233H} for the
  $624.065$\,nm line and the values by \citeads{1995A&A...296..217B}
  for the four remaining lines). In all cases we find that it is
  indeed appropriate to consider a magnetic flux in the range $\langle
  \protect {\mathrm B_{vert}} \rangle \sim 50- 200$\,G at the
  corresponding depths of line formation in the real Sun.

Photospheric Fe lines in the solar case have the property that
their strength decreases with increasing temperature in their
formation layers (see \citeads{1992oasp.book.....G}). Hence, the
hotter average stratification caused by the presence of magnetic
concentrations tends to decrease their predicted equivalent widths
  with respect to the HD case. The direct (i.e., Zeeman) effect acts
in the opposite direction. In the visible range, the former effect
dominates and systematically smaller equivalent widths result. In the
IR, Zeeman broadening is important and stronger lines (i.e.,
larger equivalent widths than in the HD case may result as a net
  effect.

A dual picture thus clearly emerges from our results. For
spectral lines in the visible wavelength range 
(Figs.~\ref{EWs_gL_0}, \ref{EWs_gL_large}, and \ref{EWs_Aspl}), Zeeman
broadening is relatively small and the dominant influence of magnetic
fields on line formation is the line-weakening effect of the
warmer average temperature structure in the revelant line-formation
layers of MHD model atmospheres (see
Fig.\ref{fig1_avTxyt_vsDepth_and_vsTau_ks1in5}). The 3D HD abundance
results will need to be amended via {\it positive} abundance
corrections to increase their value to the one appropriate to the
magnetic case considered most representative of solar
conditions.  On the other hand, Zeeman splitting becomes important in
the IR region of the spectrum and tends to be able to more than
compensate for the indirect line-weakening effect, so much so that for
lines with particularly high Zeeman sensitivity ($g_{\mathrm{L}} >
1.5$, top panels of Fig.~\ref{EWs_IR}), a net line-strengthening
effect is noticeable. The 3D HD abundance will in this case need to be
revised downward (applying {\it negative} corrections to it).

\begin{table}
  \caption{Abundance corrections for the five IR spectral lines of Fe\,{\sc i} included in
    this study.}
\label{abucorr_IR}      % is used to refer this table in the text
\centering                          % used for centering table
\begin{tabular}{c c c c c}        % centered columns (5 columns)
\hline\hline                 % inserts double horizontal lines
$\lambda$ & $\Delta \mathrm{\log} \, \epsilon \mathrm{(Fe)}_{\odot}$ & $\Delta \mathrm{\log} \, \epsilon \mathrm{(Fe)}_{\odot}$ & $\Delta \mathrm{\log} \, \epsilon \mathrm{(Fe)}_{\odot}$ & W$_{\mathrm{HD}}$ \\    % table heading
\protect  [nm]  & [dex]     &   [dex]     &   [dex]  & [pm] \\
                & ($50$\,G - HD) &  ($100$\,G - HD) & ($200$\,G - HD) & \\
\hline                        % inserts single horizontal line
 1558.826                & $ 0.00$ & $ 0.00$  &    \protect $+0.01$ & 11.63 \\
 1564.851                & $-0.05$ & $-0.07$  &    $-0.08$ &  9.57 \\
 1565.287                & $-0.03$ & $-0.05$  &    $-0.06$ &  7.18 \\
 1566.202                & $-0.02$ & $-0.02$  &    $-0.03$ & 19.88 \\
 1566.524                & $-0.01$ & $-0.01$  &    $-0.02$ &  7.67 \\
\hline                        % inserts single horizontal line
$\langle\rangle$       & $-0.022   $ &  $-0.030   $ & $-0.036   $ \\
                       & $\pm 0.017$ &  $\pm 0.026$ & $\pm 0.031$ \\
\hline\hline
\end{tabular}
\tablefoot{Same order as per columns 1-5 of Table~\ref{abucorr_visible}. In the bottom row, we provide the (unweighted) mean of the abundance corrections and
    corresponding rms.\\
}
\end{table}

   %______________________________________________
   \begin{figure}
   \centering
   \includegraphics[width=0.45\textwidth]{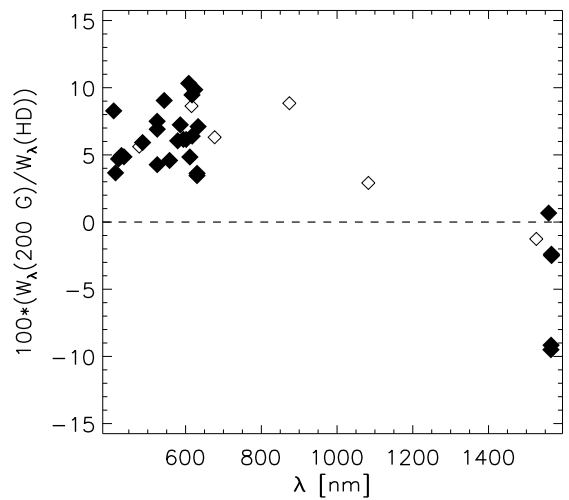}
   \caption{Equivalent width percentage change of the 200 G case compared to HD, plotted against line
   rest wavelength. Filled diamonds are for the 28 iron lines listed in Table~\ref{lines_params}, empty symbols for 6 lines of other elements which were used to complement our line list and for testing.}
   \label{EW_trends}
   \end{figure}
   %______________________________________________

   These results are in line with our previous exploratory
     findings 
     \citepads{2010ApJ...724.1536F}. A cross-check can be made since
     the three lines in that paper are included in the current study.
     The agreement is very good. Any small difference in the
     theoretical equivalent width or in the derived abundance
     corrections (generally in the sense of slightly more significant
     effects having been found now) are due to better spectral
     sampling and continuum coverage in the present study. We chose to
     decrease the number of snapshots and the spatial subsampling
     employed compared to our previous study, only after having made
     sure that this did not have a significant impact on the
     equivalent width of the lines.

   To confirm that the physical processes involved can explain the
   effects we found, we checked for any abundance correction trends 
   with the parameters characterizing the line formation,
   namely, rest wavelength of the line, lower level excitation
   potential and oscillator strength of the given transition, level
   multiplicity (2s+1) and total angular momentum J, line centre
   formation height, Land\'e factor, and HD line strength. Here we
   present a figure (Fig.~\ref{EW_trends}) corresponding to the only
   trend retrievable, namely the sign of the abundance correction
   changes as one moves from wavelengths in the visible $\lambda$
   range to the IR.  In Fig.~\ref{EW_trends} we have exceptionally
   included results for a few lines of elements other than iron.
   These are C\,{\sc i} $477.590$~nm, O\,{\sc i} $615.818$~nm,
     Ni\,{\sc i} $676.777$~nm
     (used in SOHO MDI magnetogram images),
     Mn\,{\sc i} $874.091$~nm,
     Si\,{\sc i} $1082.71$~nm (this IR line is the strongest in the
     sample we study, i.e., having an equivalent width $>500$~m\AA),
     and finally, Mn\,{\sc i} $1526.27$~nm
     (\citeads{2009ApJ...690..416A, 2009ASPC..405..215A}).
     We consider these additional lines of other elements 
     because, as shown in Fig.~\ref{EW_trends}, they cover the region of
     intermediate wavelengths in the visible to near-IR where our
     sample has no iron lines, and provide hints that magnetic flux acts
     similarly also on lines of different elements than iron.
   Interestingly, the lines of
   the other elements in fact follow a similar trend to the one seen for iron
   lines. This would seem to suggest that the effects encountered can
   be generalized to other elements according to considerations on a
   given line's formation height and temperature sensitivity, on the
   one hand, and Land\'e factor and wavelength, on the other.  We also
 see that the percentage change in equivalent width decreases smoothly
 from the lower to the upper wavelengths range limits for those lines.

 In general, we find significant differences in the effects for
 different lines. Correspondingly, the abundance corrections vary from
 quite small ($|\Delta \mathrm{\log} \, \epsilon
 \mathrm{(Fe)}_{\odot}| \lessapprox 0.02$\,dex) to significant (up to
 $|\Delta \mathrm{\log} \, \epsilon \mathrm{(Fe)}_{\odot}| \sim
 0.15$\,dex in the 200 G case) for all but a few of the spectral
 lines analysed. Lines affected to a level below $0.02$\,dex represent
 only rare exceptions.

 We focussed also on the subset of lines that we have in common with
 the absolute solar iron abundance determination of
 \citetads{2000A&A...359..743A}. These all have $g_L \ge 1$. The
 abundance corrections found here for them are among the most
 significant of those we find to be applicable to the lines we have
 studied.  Interestingly, for this subset of lines, the effects due to
 the presence of magnetic fields can be seen as linked to their
 formation height. That is, they follow the trend seen in
 Fig.~\ref{fig1_avTxyt_vsDepth_and_vsTau_ks1in5}, according to which
 lines formed between a height of $200$ and $400$~km feel increasingly
 strong indirect effects. The abundance corrections for lines in this
 subset follow the trend expected, based on their
 formation heights (see values in \citeads[][Table~2]{1989GurtKost}).
 Other spectral features used in the study by
 \citetads{2000A&A...359..743A} are also likely to be affected by
 significant abundance corrections, in particular for lines with a small
 Land\'e factor, for which the indirect effect found here can act
 undisturbed. Thus, the last word on the solar iron photospheric
   abundance has probably not yet been spoken.

   Our results also urge more investigation into the magnitude of the
   magnetic effects for other chemical elements than iron. In
   particular, this is true for those elements whose abundances have in the
   last decade undergone a revision that has caused large disagreement
   between standard solar models and helioseismology constraints.
   It is crucial that also oxygen, carbon,
   neon, nitrogen and silicon - i.e.~ those elements playing a major role
   in determining the total solar metallicity Z/X - be redetermined via
   targeted studies simultaneously including the most accurate atomic
   and observational data and non--LTE, 3D, and magnetic effects.

   Since all spectral lines have a more or less pronounced sensitivity
   to the temperature values of the layers where they are formed, and
   since direct magnetic sensitivity of the lines, if present, induces
   line broadening, one can make a broad division into the following
   expected general behaviour, depending on the chemical element and
   ionization stage considered:

   \begin{itemize}
   \item Spectral lines for which magnetic broadening acts in the
     opposite sense to the indirect temperature effect. This is the
     case of both neutral and singly-ionized iron lines in the Sun
     and of other elements having the same temperature sensitivity.
     In this category, there will be some rare instances of
     close-to-negligible net abundance corrections.
     In general, though, the direct and the indirect effect will not
     balance out exactly. The net resulting abundance correction
     can be either positive or negative, with a sign inversion
     expected when moving from UV to IR wavelengths, i.e., towards the
     spectral range where Zeeman broadening becomes important.

   \item Spectral lines for which the indirect temperature effect acts
     in the same direction as magnetic broadening. These will tend to
     become stronger in MHD simulations, thus the relevant abundance
     corrections to be applied to 3D HD abundance determinations will
     always be {\it negative}. Still, their magnitude should be
     investigated on a case-by-case basis. The net abundance
     corrections will be maximized, because of the direct and indirect
     effects working together to reinforce the spectral line. Although
     Zeeman splitting for most lines of chemical elements other than
     iron may not be as important, the spectral features in this
     category will still tend to become stronger in the warmer MHD
     temperature stratification thanks to the indirect effect. 
     One important example are lines of neutral oxygen (since that
       chemical element remains mostly neutral in the solar
       photosphere due to its much higher ionization potential
       compared to e.~g. iron), for which our preliminary abundance
       results of work included in an upcoming publication seem to
       indicate significant effects.

   \end{itemize}

   Given the differential nature of our investigation, we expect that the
   neglect of non--LTE effects on Fe\,{\sc i} lines should not change our
   conclusions significantly. In fact, if anything, non--LTE
   corrections for neutral iron are expected to push 
   the solar abundance determination even
   higher due to over-ionization caused by the near-UV radiation field, hence
   producing under-population of the neutral iron atomic levels and thus
   weaker emergent line profiles compared to LTE due to line opacity deficit
   (\citetads{2005A&A...442..643C}; \citetads{2001ApJ...550..970S}). In
   fact, attempts at improving the 3D HD LTE-based
   \citetads{2000A&A...359..743A} study also in this sense have already
   pushed the solar iron abundance back up to $\log(Fe)_{\odot} = 7.50$,
   while achieving better agreement with helioseismology, and while
   preserving the agreement between values based on lines from neutral and
   singly-ionized Fe and the small line-to-line scatter and absence of trends
   with excitation potential (see Table 1 and Fig. 6 in
   \citeads{2009ARA&A..47..481A}). Based on that result and on our current
   findings regarding the systematic abundance corrections applicable due to
   magnetic fields (with likely additional effects expected in 3D non--LTE, see \citeads{2012A&AHolzSol}), the real photospheric solar iron content may well be
   $\log(Fe)_{\odot} > 7.50$, i.e. in a relatively high range compared to the
   best estimates obtained in the past decade. A less ``anemic'' Sun would 
   better agree with new, accurate determinations of the solar Si
   content accounting for non--LTE effects \citepads{2012KPCB...28...49S}.

     Both of these changes, together with hints that
     3D-based revisions in other chemical elements may seemingly not need to
     be drastic compared to the 1D case (e.g., see the results of
     \citeads{2009MmSAI..80..643C} based on CO$^5$BOLD 3D granulation
     models), should help alleviate the discomfort that helioseismology has
     been feeling towards low values of the solar metal content. In any case,
     it will be very interesting to investigate how 3D, non--LTE, and magnetic
     effects interact, and whether there may be some amplification or
     reduction in their magnitude when looking at a fully consistent picture
     in order to carry out an absolute abundance determination.

     A further remaining uncertainty revolves around choosing the MHD
     model that is thought to best represent the real Sun, in order to
     estimate the most appropriate abundance correction to be applied
     to the abundance derived in the HD case. In particular,
       further testing of the realism of the MHD models must be
       carried out in terms of geometric configuration and depth
       profile of the magnetic field. Our non-magnetic model has
       significantly lower values of the photospheric temperature for the average stratification at $\log
  \tau_{500\,\mathrm{nm}}<0$ compared with the
       temperature structure of the semi-empirical solar atmosphere derived by        \citetads{1974SoPh...39...19H}. For those same optical depths, our MHD models are
       progressively closer to it as magnetic flux is increased, while very good agreement is generally reached for deep layers. Compared to the theoretical results of 
       \citetads{1998ApJ...499..914S} (see their Fig. 15, also shown in Fig. 14 of \citeads{2009LRSP....6....2N}), our HD atmospheric temperatures
       achieve similar to significantly better agreement (in photospheric and deep layers,
       respectively) with the stratification of \citetads{1974SoPh...39...19H}. A
     number of observational results are certainly helping to
     constrain the possibilities, and comparisons with future,
     high-quality, targeted solar data are set to improve this
     scenario even further.  In this sense, we have embarked on a
     detailed comparison and testing in terms of line parameters and
     thermodynamics, of centre-to-limb variation (CLV) and Stokes
     parameters, as well as of line fitting and absolute abundance
     derivation, but this goes beyond the scope of the present paper and
     is part of work in progress. As an example, the tests we
       carried out and discuss in a separate article (Beck et al.
       2012, {\it submitted}), % A&A is able to deal with upcoming publications
       show that the agreement with the observational data we studied of predicted properties (for spatially- and spectrally-degraded spectral lines),
       such as the rms fluctuations at different line depression levels of bisector position/velocity, bisector intensity, and line width, may at least in some cases improve when including magnetic fields in the
       simulations. It will also be important to continue along the
     line traced by initial efforts of using Stokes spectra from
     magneto-convection simulations (\citeads{2011SPD....42.0804S}) 
     to make comparison possible of theoretical predictions with the best available current and
     future spectropolarimetric observational data.

%______________________________________________________________

\section{Conclusions}
\label{section:conclusions}

We have demonstrated that appreciable effects on a set of iron
spectral lines are found when introducing magnetic fields in
3D, time-dependent radiation-hydrodynamics simulations
of solar surface convection. Even the most advanced chemical
composition studies so far, using 3D convection modelling, may
  still be affected in a significant way in terms of metallicity
  determinations owing to their non-magnetic assumption. We here
determined the associated abundance corrections to be applied when
abundances are derived via a purely HD approach.
In the following, we briefly highlight our main findings.
 
We have presented evidence that the magnetic field causes
both Zeeman broadening in the lines with non-zero Land\'e factors and also
     changes in the average temperature stratification of the cells.
     Since the majority of the lines have non-zero Land\'e factors and
     are also temperature-sensitive to some extent, no spectral line is in
     principle exempt from relevant abundance corrections. The effects are
     therefore ubiquitous.
     The direct (line-strengthening) effect of magnetic field on line
     formation via Zeeman broadening is generally (except in the IR) less
     important in magnitude
     than the indirect effect of temperature stratification changes.

The abundance corrections are approximately proportional to the
     average unsigned magnetic field strength $\langle \mathrm{|B_{vert}|} \rangle$.
     For $\langle \mathrm{B_{vert}} \rangle=100$ G, the abundance corrections
     are $|\Delta \log(Fe)| \sim 0.04$\,dex for our full selection of
     iron spectral lines in the visible wavelength range and 
     $|\Delta \log(Fe)| \sim 0.07$\,dex 
     for the subgroup in common with \citetads{2000A&A...359..743A}.

If, as a number of recent results seem to suggest, the
     ``quiet'' solar photosphere is in fact teeming with hidden
     magnetic flux, the abundance of iron based on HD modelling will need to be revised
     upwards. It is likely that a more accurate
     estimate for the solar iron content may be $\log(Fe) \gtrsim 7.50$.

\begin{acknowledgements}
  We gratefully acknowledge financial support by the European
  Commission through the SOLAIRE Network (MTRN-CT-2006-035484), by the
  Spanish Ministry of Research and Innovation through projects
  AYA2007-66502, CSD2007-00050, AYA2007-63881, and AYA2011-24808, and
  by the Danish Research Agency - Danish Natural Science Research
  Council. We acknowledge the computing time granted through the DEISA
SolarAct and PRACE SunFlare projects and the corresponding use of the HLRS and FZJ-JSC JUROPA supercomputer installations, as well as the computing
time allocated via RES calls at the MareNostrum (BSC/CNS, Spain)
and LaPalma (IAC/RES, Spain) supercomputers. We also appreciate the
use of the facilities at the Danish Center for Scientific Computing (DCSC-KU,
Denmark). We thank the referee
%, M. Steffen, 
for helpful comments and A. de Vicente (IAC Condor
  management), H.  Socas-Navarro, N. Shchukina, J. Trujillo Bueno, C.
  Beck, C.  Allende-Prieto, and J.~M.  Borrero for fruitful
  interaction.
\end{acknowledgements}

%%% If using natbib (BibTex), i.e. entries directly formatted from
%%% ADS.  This uses file .bib specified in \bibliography command.
%
\bibliographystyle{aa}    %% bibliography style file aa.bst from A&A
\bibliography{Fabbian+2012_Fe_LTE_Sun_MHD_LE}    %% .bib which contains Bibtex entries copied from ADS

\newcommand{\noop}[1]{}
\begin{thebibliography}{64}
\expandafter\ifx\csname natexlab\endcsname\relax\def\natexlab#1{#1}\fi

\bibitem[{{Antia} \& {Basu}(2005)}]{2005ApJ...620L.129A}
{Antia}, H.~M. \& {Basu}, S. 2005, \apjl, 620, L129

\bibitem[{{Asensio Ramos}(2009)}]{2009ApJ...690..416A}
{Asensio Ramos}, A. 2009, \apj, 690, 416

\bibitem[{{Asensio Ramos} {et~al.}(2009){Asensio Ramos}, {Mart{\'{\i}}nez
  Gonz{\'a}lez}, {L{\'o}pez Ariste}, {Trujillo Bueno}, \&
  {Collados}}]{2009ASPC..405..215A}
{Asensio Ramos}, A., {Mart{\'{\i}}nez Gonz{\'a}lez}, M.~J., {L{\'o}pez Ariste},
  A., {Trujillo Bueno}, J., \& {Collados}, M. 2009, in Astronomical Society of
  the Pacific Conference Series, Vol. 405, Solar Polarization 5: In Honor of
  Jan Stenflo, ed. {S.~V.~Berdyugina, K.~N.~Nagendra, \& R.~Ramelli}, 215

\bibitem[{{Asplund}(2005)}]{2005ARA&A..43..481A}
{Asplund}, M. 2005, \araa, 43, 481

\bibitem[{{Asplund} {et~al.}(2009){Asplund}, {Grevesse}, {Sauval}, \&
  {Scott}}]{2009ARA&A..47..481A}
{Asplund}, M., {Grevesse}, N., {Sauval}, A.~J., \& {Scott}, P. 2009, \araa, 47,
  481

\bibitem[{{Asplund} {et~al.}(2000{\natexlab{a}}){Asplund}, {Nordlund},
  {Trampedach}, {Allende Prieto}, \& {Stein}}]{2000A&A...359..729A}
{Asplund}, M., {Nordlund}, {\AA}., {Trampedach}, R., {Allende Prieto}, C., \&
  {Stein}, R.~F. 2000{\natexlab{a}}, \aap, 359, 729

\bibitem[{{Asplund} {et~al.}(2000{\natexlab{b}}){Asplund}, {Nordlund},
  {Trampedach}, \& {Stein}}]{2000A&A...359..743A}
{Asplund}, M., {Nordlund}, {\AA}., {Trampedach}, R., \& {Stein}, R.~F.
  2000{\natexlab{b}}, \aap, 359, 743

\bibitem[{{Ayres}(2007)}]{2007AAS...211.5908A}
{Ayres}, T.~R. 2007, in Bulletin of the American Astronomical Society, Vol.~38,
  American Astronomical Society Meeting Abstracts, 842

\bibitem[{{Ayres}(2008)}]{2008ApJ...686..731A}
{Ayres}, T.~R. 2008, \apj, 686, 731

\bibitem[{{Ayres}(2012)}]{2012AAS...21914408A}
{Ayres}, T.~R. 2012, in American Astronomical Society Meeting Abstracts, Vol.
  219, American Astronomical Society Meeting Abstracts, \#144.08

\bibitem[{{Ayres} {et~al.}(2006){Ayres}, {Plymate}, \&
  {Keller}}]{2006ApJS..165..618A}
{Ayres}, T.~R., {Plymate}, C., \& {Keller}, C.~U. 2006, \apjs, 165, 618

\bibitem[{{Barthol} {et~al.}(2011){Barthol}, {Gandorfer}, {Solanki},
  {Sch{\"u}ssler}, {Chares}, {Curdt}, {Deutsch}, {Feller}, {Germerott},
  {Grauf}, {Heerlein}, {Hirzberger}, {Kolleck}, {Meller}, {M{\"u}ller},
  {Riethm{\"u}ller}, {Tomasch}, {Kn{\"o}lker}, {Lites}, {Card}, {Elmore},
  {Fox}, {Lecinski}, {Nelson}, {Summers}, {Watt}, {Mart{\'{\i}}nez Pillet},
  {Bonet}, {Schmidt}, {Berkefeld}, {Title}, {Domingo}, {Gasent Blesa}, {Del
  Toro Iniesta}, {L{\'o}pez Jim{\'e}nez}, {{\'A}lvarez-Herrero},
  {Sabau-Graziati}, {Widani}, {Haberler}, {H{\"a}rtel}, {Kampf}, {Levin},
  {P{\'e}rez Grande}, {Sanz-Andr{\'e}s}, \& {Schmidt}}]{2011SoPh..268....1B}
{Barthol}, P., {Gandorfer}, A., {Solanki}, S.~K., {et~al.} 2011, \solphys, 268,
  1

\bibitem[{{Beeck} {et~al.}(2012){Beeck}, {Collet}, {Steffen}, {Asplund},
  {Cameron}, {Freytag}, {Hayek}, {Ludwig}, \&
  {Sch{\"u}ssler}}]{2012A&A...539A.121B}
{Beeck}, B., {Collet}, R., {Steffen}, M., {et~al.} 2012, \aap, 539, A121

\bibitem[{{Berger} {et~al.}(2004){Berger}, {Rouppe van der Voort},
  {L{\"o}fdahl}, {Carlsson}, {Fossum}, {Hansteen}, {Marthinussen}, {Title}, \&
  {Scharmer}}]{2004A&A...428..613B}
{Berger}, T.~E., {Rouppe van der Voort}, L.~H.~M., {L{\"o}fdahl}, M.~G.,
  {et~al.} 2004, \aap, 428, 613

\bibitem[{{Blackwell} {et~al.}(1986){Blackwell}, {Booth}, {Haddock}, {Petford},
  \& {Leggett}}]{1986MNRAS.220..549B}
{Blackwell}, D.~E., {Booth}, A.~J., {Haddock}, D.~J., {Petford}, A.~D., \&
  {Leggett}, S.~K. 1986, \mnras, 220, 549

\bibitem[{{Blackwell} {et~al.}(1984){Blackwell}, {Booth}, \&
  {Petford}}]{1984A&A...132..236B}
{Blackwell}, D.~E., {Booth}, A.~J., \& {Petford}, A.~D. 1984, \aap, 132, 236

\bibitem[{{Blackwell} {et~al.}(1995){Blackwell}, {Lynas-Gray}, \&
  {Smith}}]{1995A&A...296..217B}
{Blackwell}, D.~E., {Lynas-Gray}, A.~E., \& {Smith}, G. 1995, \aap, 296, 217

\bibitem[{{Bonet} {et~al.}(2010){Bonet}, {M{\'a}rquez}, {S{\'a}nchez Almeida},
  {Palacios}, {Mart{\'{\i}}nez Pillet}, {Solanki}, {del Toro Iniesta},
  {Domingo}, {Berkefeld}, {Schmidt}, {Gandorfer}, {Barthol}, \&
  {Kn{\"o}lker}}]{2010ApJ...723L.139B}
{Bonet}, J.~A., {M{\'a}rquez}, I., {S{\'a}nchez Almeida}, J., {et~al.} 2010,
  \apjl, 723, L139

\bibitem[{{Borrero}(2008)}]{Borrero2008}
{Borrero}, J.~M. 2008, \apj, 673, 470

\bibitem[{{Caffau} {et~al.}(2010){Caffau}, {Ludwig}, {Bonifacio}, {Faraggiana},
  {Steffen}, {Freytag}, {Kamp}, \& {Ayres}}]{2010A&A...514A..92C}
{Caffau}, E., {Ludwig}, H.-G., {Bonifacio}, P., {et~al.} 2010, \aap, 514, A92

\bibitem[{{Caffau} {et~al.}(2009){Caffau}, {Ludwig}, \&
  {Steffen}}]{2009MmSAI..80..643C}
{Caffau}, E., {Ludwig}, H.-G., \& {Steffen}, M. 2009, \memsai, 80, 643

\bibitem[{{Caffau} {et~al.}(2008){Caffau}, {Ludwig}, {Steffen}, {Ayres},
  {Bonifacio}, {Cayrel}, {Freytag}, \& {Plez}}]{2008A&A...488.1031C}
{Caffau}, E., {Ludwig}, H.-G., {Steffen}, M., {et~al.} 2008, \aap, 488, 1031

\bibitem[{{Caffau} {et~al.}(2011){Caffau}, {Ludwig}, {Steffen}, {Freytag}, \&
  {Bonifacio}}]{2011SoPh..268..255C}
{Caffau}, E., {Ludwig}, H.-G., {Steffen}, M., {Freytag}, B., \& {Bonifacio}, P.
  2011, \solphys, 268, 255

\bibitem[{{Caffau} {et~al.}(2007){Caffau}, {Steffen}, {Sbordone}, {Ludwig}, \&
  {Bonifacio}}]{2007A&A...473L...9C}
{Caffau}, E., {Steffen}, M., {Sbordone}, L., {Ludwig}, H.-G., \& {Bonifacio},
  P. 2007, \aap, 473, L9

\bibitem[{Cattaneo {et~al.}(2003)Cattaneo, Emonet, \&
  Weiss}]{CattEmonWeiss2003}
Cattaneo, F., Emonet, T., \& Weiss, N. 2003, Astrophys. J., 588, 1183

\bibitem[{{Cheung} {et~al.}(2008){Cheung}, {Sch{\"u}ssler}, {Tarbell}, \&
  {Title}}]{2008ApJ...687.1373C}
{Cheung}, M.~C.~M., {Sch{\"u}ssler}, M., {Tarbell}, T.~D., \& {Title}, A.~M.
  2008, \apj, 687, 1373

\bibitem[{{Collet} {et~al.}(2005){Collet}, {Asplund}, \&
  {Th{\'e}venin}}]{2005A&A...442..643C}
{Collet}, R., {Asplund}, M., \& {Th{\'e}venin}, F. 2005, \aap, 442, 643

\bibitem[{{Delahaye} \& {Pinsonneault}(2006)}]{2006ApJ...649..529D}
{Delahaye}, F. \& {Pinsonneault}, M.~H. 2006, \apj, 649, 529

\bibitem[{{Fabbian} {et~al.}(2010){Fabbian}, {Khomenko}, {Moreno-Insertis}, \&
  {Nordlund}}]{2010ApJ...724.1536F}
{Fabbian}, D., {Khomenko}, E., {Moreno-Insertis}, F., \& {Nordlund}, {\AA}.
  2010, \apj, 724, 1536

\bibitem[{{Galsgaard} \& {Nordlund}(1996)}]{1996JGR...10113445G}
{Galsgaard}, K. \& {Nordlund}, {\AA}. 1996, \jgr, 101, 13445

\bibitem[{{Gray}(1992)}]{1992oasp.book.....G}
{Gray}, D.~F. 1992, {The observation and analysis of stellar photospheres.},
  ed. {Camb.~Astrophys.~Ser., Vol.~20}

\bibitem[{{Gurtovenko} \& {Kostik}(1989)}]{1989GurtKost}
{Gurtovenko}, E.~A. \& {Kostik}, R.~I. 1989, {Fraunhofer Spectrum and the
  System of Solar Oscillator Strengths}, ed. {Naukova Dumka, Kiev}

\bibitem[{{Holweger} {et~al.}(1990){Holweger}, {Heise}, \&
  {Kock}}]{1990A&A...232..510H}
{Holweger}, H., {Heise}, C., \& {Kock}, M. 1990, \aap, 232, 510

\bibitem[{{Holweger} {et~al.}(1995){Holweger}, {Kock}, \&
  {Bard}}]{1995A&A...296..233H}
{Holweger}, H., {Kock}, M., \& {Bard}, A. 1995, \aap, 296, 233

\bibitem[{{Holweger} \& {Mueller}(1974)}]{1974SoPh...39...19H}
{Holweger}, H. \& {Mueller}, E.~A. 1974, \solphys, 39, 19

\bibitem[{{Holzreuter} \& {Solanki}(2012)}]{2012A&AHolzSol}
{Holzreuter}, R. \& {Solanki}, S.~K. 2012, \aap, ???

\bibitem[{{Lodders} {et~al.}(2009){Lodders}, {Palme}, \&
  {Gail}}]{2009LanB...4B...44L}
{Lodders}, K., {Palme}, H., \& {Gail}, H.-P. 2009, in ''Landolt-B{\"o}rnstein -
  Group VI Astronomy and Astrophysics Numerical Data and Functional
  Relationships in Science and Technology Volume, ed. {J.~E.~Tr{\"u}mper}, 44

\bibitem[{{Maltby} {et~al.}(1986){Maltby}, {Avrett}, {Carlsson},
  {Kjeldseth-Moe}, {Kurucz}, \& {Loeser}}]{1986ApJ...306..284M}
{Maltby}, P., {Avrett}, E.~H., {Carlsson}, M., {et~al.} 1986, \apj, 306, 284

\bibitem[{{Mart{\'{\i}}nez Pillet} {et~al.}(2011){Mart{\'{\i}}nez Pillet}, {Del
  Toro Iniesta}, {{\'A}lvarez-Herrero}, {Domingo}, {Bonet}, {Gonz{\'a}lez
  Fern{\'a}ndez}, {L{\'o}pez Jim{\'e}nez}, {Pastor}, {Gasent Blesa}, {Mellado},
  {Piqueras}, {Aparicio}, {Balaguer}, {Ballesteros}, {Belenguer}, {Bellot
  Rubio}, {Berkefeld}, {Collados}, {Deutsch}, {Feller}, {Girela}, {Grauf},
  {Heredero}, {Herranz}, {Jer{\'o}nimo}, {Laguna}, {Meller}, {Men{\'e}ndez},
  {Morales}, {Orozco Su{\'a}rez}, {Ramos}, {Reina}, {Ramos},
  {Rodr{\'{\i}}guez}, {S{\'a}nchez}, {Uribe-Patarroyo}, {Barthol}, {Gandorfer},
  {Knoelker}, {Schmidt}, {Solanki}, \& {Vargas
  Dom{\'{\i}}nguez}}]{2011SoPh..268...57M}
{Mart{\'{\i}}nez Pillet}, V., {Del Toro Iniesta}, J.~C., {{\'A}lvarez-Herrero},
  A., {et~al.} 2011, \solphys, 268, 57

\bibitem[{{Mel{\'e}ndez} \& {Barbuy}(2009)}]{2009A&A...497..611M}
{Mel{\'e}ndez}, J. \& {Barbuy}, B. 2009, \aap, 497, 611

\bibitem[{{Muller}(1977)}]{1977SoPh...52..249M}
{Muller}, R. 1977, \solphys, 52, 249

\bibitem[{{Narayan} \& {Scharmer}(2010)}]{2010A&A...524A...3N}
{Narayan}, G. \& {Scharmer}, G.~B. 2010, \aap, 524, A3

\bibitem[{{Neckel} \& {Labs}(1984)}]{1984SoPh...90..205N}
{Neckel}, H. \& {Labs}, D. 1984, \solphys, 90, 205

\bibitem[{{Nordlund}(1984)}]{Nordlund1984}
{Nordlund}, A. 1984, in ESA Special Publication, Vol. 220, ESA Special
  Publication, ed. {T.~D.~Guyenne \& J.~J.~Hunt}, 37--46

\bibitem[{{Nordlund} {et~al.}(2009){Nordlund}, {Stein}, \&
  {Asplund}}]{2009LRSP....6....2N}
{Nordlund}, {\AA}., {Stein}, R.~F., \& {Asplund}, M. 2009, Living Reviews in
  Solar Physics, 6, 2

\bibitem[{{S{\'a}nchez Almeida} \& {Mart{\'{\i}}nez
  Gonz{\'a}lez}(2011)}]{2011ASPC..437..451S}
{S{\'a}nchez Almeida}, J. \& {Mart{\'{\i}}nez Gonz{\'a}lez}, M. 2011, in
  Astronomical Society of the Pacific Conference Series, Vol. 437, Solar
  Polarization 6, ed. {J.~R.~Kuhn, D.~M.~Harrington, H.~Lin, S.~V.~Berdyugina,
  J.~Trujillo-Bueno, S.~L.~Keil, \& T.~Rimmele}, 451

\bibitem[{{Schmidt} {et~al.}(1988){Schmidt}, {Grossmann-Doerth}, \&
  {Schroeter}}]{1988A&A...197..306S}
{Schmidt}, W., {Grossmann-Doerth}, U., \& {Schroeter}, E.~H. 1988, \aap, 197,
  306

\bibitem[{{Schuessler} \& {Solanki}(1988)}]{SchussSolanki1988}
{Schuessler}, M. \& {Solanki}, S.~K. 1988, \aap, 192, 338

\bibitem[{{Sch{\"u}ssler} \& {V{\"o}gler}(2006)}]{2006ApJ...641L..73S}
{Sch{\"u}ssler}, M. \& {V{\"o}gler}, A. 2006, \apjl, 641, L73

\bibitem[{{Shchukina} \& {Trujillo Bueno}(2001)}]{2001ApJ...550..970S}
{Shchukina}, N. \& {Trujillo Bueno}, J. 2001, \apj, 550, 970

\bibitem[{{Shchukina} \& {Sukhorukov}(2012)}]{2012KPCB...28...49S}
{Shchukina}, N.~G. \& {Sukhorukov}, A.~V. 2012, Kinematics and Physics of
  Celestial Bodies, 28, 49

\bibitem[{{Socas-Navarro}(2001)}]{2001ASPC..236..487S}
{Socas-Navarro}, H. 2001, in Astronomical Society of the Pacific Conference
  Series, Vol. 236, Advanced Solar Polarimetry -- Theory, Observation, and
  Instrumentation, ed. {M.~Sigwarth}, 487

\bibitem[{{Solanki} {et~al.}(2010){Solanki}, {Barthol}, {Danilovic}, {Feller},
  {Gandorfer}, {Hirzberger}, {Riethm{\"u}ller}, {Sch{\"u}ssler}, {Bonet},
  {Mart{\'{\i}}nez Pillet}, {del Toro Iniesta}, {Domingo}, {Palacios},
  {Kn{\"o}lker}, {Bello Gonz{\'a}lez}, {Berkefeld}, {Franz}, {Schmidt}, \&
  {Title}}]{2010ApJ...723L.127S}
{Solanki}, S.~K., {Barthol}, P., {Danilovic}, S., {et~al.} 2010, \apjl, 723,
  L127

\bibitem[{{Spruit}(1976)}]{Spruit1976}
{Spruit}, H.~C. 1976, \solphys, 50, 269

\bibitem[{{Stein} {et~al.}(2011){Stein}, {Georgobiani}, {Nordlund}, \&
  {Lagerfjard}}]{2011SPD....42.0804S}
{Stein}, R.~F., {Georgobiani}, D., {Nordlund}, A., \& {Lagerfjard}, A. 2011, in
  AAS/Solar Physics Division Abstracts \#42, 804

\bibitem[{{Stein} \& {Nordlund}(1998)}]{1998ApJ...499..914S}
{Stein}, R.~F. \& {Nordlund}, A. 1998, \apj, 499, 914

\bibitem[{{Stein} \& {Nordlund}(2000)}]{2000SoPh..192...91S}
{Stein}, R.~F. \& {Nordlund}, {\AA}. 2000, \solphys, 192, 91

\bibitem[{{Stenflo}(1973)}]{Stenflo1973}
{Stenflo}, J.~O. 1973, \solphys, 32, 41

\bibitem[{{Title} {et~al.}(1992){Title}, {Topka}, {Tarbell}, {Schmidt},
  {Balke}, \& {Scharmer}}]{1992ApJ...393..782T}
{Title}, A.~M., {Topka}, K.~P., {Tarbell}, T.~D., {et~al.} 1992, \apj, 393, 782

\bibitem[{{Trujillo Bueno} \& {Shchukina}(2009)}]{2009ApJ...694.1364T}
{Trujillo Bueno}, J. \& {Shchukina}, N. 2009, \apj, 694, 1364

\bibitem[{{Trujillo Bueno} {et~al.}(2004){Trujillo Bueno}, {Shchukina}, \&
  {Asensio Ramos}}]{2004Natur.430..326T}
{Trujillo Bueno}, J., {Shchukina}, N., \& {Asensio Ramos}, A. 2004, \nat, 430,
  326

\bibitem[{{Unsold}(1955)}]{1955QB461.U55......}
{Unsold}, A. 1955, {Physik der Sternatmospharen, MIT besonderer
  Berucksichtigung der Sonne.}, ed. {Unsold, A.}

\bibitem[{V{\"{o}}gler(2005)}]{Vogler2005}
V{\"{o}}gler, A. 2005, Mem. Soc. Astron. Ital., 76, 842

\bibitem[{{V{\"o}gler} {et~al.}(2005){V{\"o}gler}, {Shelyag}, {Sch{\"u}ssler},
  {Cattaneo}, {Emonet}, \& {Linde}}]{2005A&A...429..335V}
{V{\"o}gler}, A., {Shelyag}, S., {Sch{\"u}ssler}, M., {et~al.} 2005, \aap, 429,
  335

\end{thebibliography}
%
%%%

\end{document}